\documentclass[aps,prd,twocolumn,longbibliography,preprintnumbers,superscriptaddress,nofootinbib,floatfix]{revtex4-1}

\usepackage{amsmath}
\usepackage{bm}
\usepackage{amsfonts}
\usepackage{graphicx}
\usepackage{epstopdf}
\usepackage{array}
\usepackage{soul}
\usepackage{color}
\usepackage{hyperref}
\usepackage{amssymb}

\newcommand{\ket}[1]{| #1 \rangle}
\newcommand{\bra}[1]{\langle #1 |}

\hypersetup{
    colorlinks=true,
    urlcolor=blue,
    citecolor=blue,
    linkcolor=blue,
    menucolor=blue,
    anchorcolor=blue,
    filecolor=blue
}

\everymath{\displaystyle}
\smallskipamount

\newcommand{\FirstAffiliation}{\affiliation{
	Arnold Sommerfeld Center,
	Ludwig-Maximilians-Universit{\"a}t,
	Theresienstra{\ss}e 37,
	80333 M{\"u}nchen,
	Germany}}
 
\newcommand{\SecondAffiliation}{\affiliation{
	Max-Planck-Institut f{\"u}r Physik,
   Boltzmannstrasse 8,
85748 Garching,
	Germany}}

\newcommand{\ThirdAffiliation}{\affiliation{
    Tsung-Dao Lee Institute and School of Physics and Astronomy,
    Shanghai Jiao Tong University, Shengrong Road 520, 201210
    Shanghai, China}}

\newcommand{\FifthAffiliation}{\affiliation{Institut de F\'isica d’Altes Energies (IFAE) and The Barcelona Institute of Science and Technology (BIST), Campus UAB, 08193 Bellaterra (Barcelona), Spain
 }}
\usepackage{tikz,xcolor}
\definecolor{lime}{HTML}{A6CE39}
\DeclareRobustCommand{\orcidicon}{
	\begin{tikzpicture}
	\draw[lime, fill=lime] (0,0) 
	circle [radius=0.16] 
	node[white] {{\fontfamily{qag}\selectfont \tiny ID}};
	\draw[white, fill=white] (-0.0625,0.095) 
	circle [radius=0.007];
	\end{tikzpicture}
	\hspace{-2mm}
}
\foreach \x in {A, ..., Z}{
	\expandafter\xdef\csname orcid\x\endcsname{\noexpand\href{https://orcid.org/\csname orcidauthor\x\endcsname}{\noexpand\orcidicon}}
}

\begin{document}


\title{Memory Burden Effect in Black Holes and Solitons:\\ Implications for PBH}

\author{Gia Dvali}
\FirstAffiliation
\SecondAffiliation

\author{Juan Sebasti{\'a}n Valbuena-Berm{\'u}dez\!\orcidA{}} 
\email{juanvalbuena@ifae.es}
\FifthAffiliation

\author{Michael Zantedeschi\!\orcidB{}}
\email{zantedeschim@sjtu.edu.cn}
\ThirdAffiliation

\begin{abstract}
 The essence of the \textit{memory burden} effect is that a load of information carried by a system stabilizes it. 
This universal effect is especially prominent in systems with a high capacity of information storage, such as black holes and other objects with maximal microstate degeneracy, the entities universally referred to as \textit{saturons}. 
The phenomenon has several implications. The memory burden effect suppresses a further decay of a black hole, the latest, after it has emitted about half of its initial mass. As a consequence, the light primordial black holes (PBHs), that previously were assumed to be fully evaporated, are expected to be present as viable dark matter candidates. In the present paper, we deepen the understanding of the memory burden effect.  We first identify various memory burden regimes in generic Hamiltonian systems and then establish a precise correspondence in solitons and in black holes. We make transparent, at a microscopic level, the fundamental differences between the stabilization by a quantum memory burden versus the stabilization by a long-range classical hair due to a spin or an electric charge. We identify certain new features of potential observational interest, such as the model-independent spread of the stabilized masses of initially degenerate PBHs. 
\end{abstract}
\maketitle
\section{Introduction} 

The phenomenon of memory burden, originally described in \cite{Dvali:2018xpy},  is summarized in the following statement:  \\

{\it  Information loaded in a system resists its decay.} \\

Naturally, the effect is especially sound in systems with enhanced information storage capacity. This capacity can be quantified by the number and degeneracy of microstates that the system possesses for the given values of macroscopic parameters, such as, e.g., the radius $R$ and the total energy.  
  
Black holes are the prominent representatives of this category. This is obvious from Bekenstein-Hawking entropy~\cite{Bekenstein:1973ur}, 
\begin{equation} \label{SBH1} 
   S_{\rm BH} \, = \, \pi R^2 M_{\rm P}^2 \,, 
\end{equation} 
where $M_{\rm P}$ is the Planck mass. Correspondingly, it was suggested in~\cite{Dvali:2018xpy, Dvali:2020wft} that the phenomenon of memory burden must be applicable to black holes. 

This effect explains, in terms of an explicit microscopic mechanism, why, at the early stages of Hawking's decay, the information stored in a black hole cannot be released together with radiation.
 This matches the semi-classical expectation.\\
    
However, an important new feature emerges. The internally maintained information back-reacts and creates resistance against the decay of a black hole.  
This is the effect of the memory burden phenomenon in a black hole. \\

Furthermore, by performing a detailed analysis of the prototype systems, it was concluded in~\cite{Dvali:2020wft} that not only do black holes undergo the memory burden effect, but they likely are stabilized by it. 
That is, in the process of a black hole decay, the memory burden grows, and after a certain characteristic time, $t_{\rm M}$,  reaches a critical value. 
$t_{\rm M}$ is bounded from above by the age of a black hole that lost about half of its initial mass. \\
   
At this point, the black hole has evolved into a ``remnant" that cannot continue an ordinary quantum decay.  A remarkable feature is that the remnant is macroscopic, with a mass comparable to the initial black hole.
The fate of this object cannot be determined by applying a standard semi-classical analysis that would be valid for a young black hole.  \\

As pointed out in~\cite{Dvali:2020wft}, at the current level of understanding, it is not excluded that some new collective (classical) instability sets in, leading to a disintegration of the memory-burdened remnant. \\

Putting this possibility aside, after a black hole enters the memory burden phase, its evaporation must slow down dramatically. From the analyticity considerations, it was suggested in~\cite{Dvali:2020wft} that the remaining lifetime of a black hole scales as, 
  \begin{equation} \label{tauBH}
  \tau \, \sim \,   R\, S_{\rm BH}^{1+k}\,. 
  \end{equation}
 where $S_{\rm BH}$ is the entropy of initial black hole and $k>0$ is an integer.  
 
This has a number of implications which will be discussed later, after we introduce the second part of the story.  \\ 
    
On the other hand, it was shown recently~\cite{Dvali:2020wqi, Dvali:2019jjw, Dvali:2019ulr} that black holes are not the only objects with maximal information storage capacity. 
Rather, in various consistent quantum field theories (QFTs) there exist a whole class of objects, named ``saturons"~\cite{Dvali:2020wqi}, that exhibit the identical properties.  \\
 
An important aspect is that saturons can emerge in the form of solitons and other bound states in renormalizable  QFTs at weak coupling where their properties are fully under control and calculable~\cite{Dvali:2019jjw, Dvali:2019ulr, Dvali:2020wqi,  Dvali:2021jto,  Dvali:2021ooc, Dvali:2021rlf,  Dvali:2021tez, Dvali:2021ofp, Dvali:2023qlk}.  \\
   
In order to fix the definition:  a saturon represents an object that saturates the QFT upper bound on the microstate degeneracy. The bound has been formulated in~\cite{Dvali:2020wqi} and can be given in two equivalent forms as the bound on the microstate entropy, $S \equiv \ln(n_{\rm st})$, with $n_{\rm st}$ number of degenerate microstates.  \\   
  
First,  for a bound state of radius $R$ formed by QFT degrees of freedom interacting  via a running coupling $\alpha$, the upper bound on the microstate entropy is, 
       \begin{equation} \label{AlphaSbound}
       S \leqslant \frac{1}{\alpha} \,,
       \end{equation} 
 where $\alpha$ must be evaluated at the scale $1/R$. 
 
Equivalently, the bound can be written in terms of the Goldstone scale, $f$, of the spontaneously broken Poincare symmetry: 
       \begin{equation} \label{AreaSbound}
       S \leqslant\pi R^2 f^2 \,.
       \end{equation} 
As shown in~\cite{Dvali:2020wqi}, the above bounds set the maximal degeneracy reachable within the validity of the QFT description. In particular, their saturation is correlated with the saturation of unitarity by the scattering amplitudes. \\
            
As already discussed in~\cite{Dvali:2020wqi}, Bekenstein-Hawking entropy of a black hole (\ref{SBH1}) represents a particular case of saturation of both bounds. First, applied to a black hole,  the formula (\ref{AreaSbound}) is identical to (\ref{SBH1}), since for a black hole of arbitrary mass, the scale of Poincare Goldstone is given by the Planck mass, $f = M_{\rm P}$.  
  
Simultaneously, the black hole entropy (\ref{SBH1}) is also equal to (\ref{AlphaSbound}), since the gravitational coupling, evaluated at energy scale $1/R$, is $\alpha_{\rm gr} = 1/(\pi R^2M_{\rm P}^2)$.  \\
        
It has been observed that striking similarities between black holes and saturons of renormalizable QFTs extend to their other key properties:  
\begin{itemize}
    \item  Impossibility of the information-retrieval classically~\cite{Dvali:2019jjw, Dvali:2019ulr, Dvali:2020wqi, 
Dvali:2021jto, Dvali:2021rlf,  Dvali:2021tez}; 
    \item The minimal time-scale of the {\it start} of the information-retrieval, $t \sim S \, R$, which is identical to the Page's time in black holes~\cite{Dvali:2020wqi, Dvali:2021rlf,  Dvali:2021tez};
    \item  The existence of information horizon in the semi-classical theory~\cite{Dvali:2021tez}; 
    \item  Thermal-like evaporation at initial stages of the decay~\cite{Dvali:2021rlf,  Dvali:2021tez};
    \item  The relation between the maximal spin and the entropy~\cite{Dvali:2021ofp, Dvali:2023qlk}.   
\end{itemize}
The above correspondence makes the study of saturons important due to the following reasons. First, it shows that the black hole properties are not specific to gravity and can be understood within calculability domains of renormalizable  QFTs. 
   
Secondly, saturons can serve as laboratories for understanding the microscopic nature of known black hole properties and for discovering new features. \\
 
In the present paper, we shall apply this strategy to the memory burden effect in two directions. 
First, following~\cite{Dvali:2018xpy, Dvali:2020wft}, we outline the generic properties of the memory burden phenomenon within Hamiltonian models of enhanced information capacity. 
We then demonstrate the concrete manifestations of these properties within solitons, expanding the earlier analysis of~\cite{Dvali:2021tez, Valbuena:2023}. 
Next, we establish the link with analogous properties in black holes. 

In studying the memory burden effect in black holes, we shall use two strategies. 
On one hand, we shall extract the key features of the effect 
relying solely on the universality of the phenomenon and the requirement 
of QFT consistency of the system. 

On the other hand, we shall cross-check our conclusions 
with a microscopic theory of black hole's quantum $N$-portrait~\cite{Dvali:2011aa, Dvali:2012en, Dvali:2013eja}. In this picture
a black hole is described as a saturated coherent state 
of gravitons, at criticality, which makes the origin
of its enhanced information storage capacity very explicit.  

This approach allows us to predict certain new features of memory burden in black holes, moving forward from the previous studies. 

The memory burden effect has a wide range of applications. As already put forward in~\cite{Dvali:2020wft}, one immediate application is the opening up of a new window of the primordial black hole dark matter with masses below $10^{14}\text{g}$. 
In the standard treatment, this interval of masses was ignored based on the assumption of the validity of the semi-classical picture during the black hole's entire lifetime. The memory burden effect invalidates this assumption.

Some viable examples of PBH dark matter in the new mass window were given already in~\cite{Dvali:2020wft}.  
Further studies, focusing on the formation mechanisms~\cite{Dvali:2021byy} and various constraints on memory burdened PBH dark matter~\cite{Alexandre:2024nuo, Thoss:2024hsr, Balaji:2024hpu, Haque:2024eyh}, have also been performed. 

In the present paper, we shall predict a new feature of potential observational significance: the model-independent spread in PBH masses induced by the memory burden effect.   

As suggested in~\cite{Dvali:2018ytn}, another entity, likely subjected to the memory burden effect, is a de Sitter Hubble patch of radius $R$. Similarly to a black hole, de Sitter carries a maximal microstate entropy of Gibbons-Hawking~\cite{Gibbons:1977mu} given by the expression (\ref{SBH1}). 
Due to this, it falls in the category of systems with enhanced capacity for information storage. This creates an avenue for studying the imprints of the memory burden effect in inflationary cosmology, with the first steps taken in~\cite{Dvali:2018ytn}. \\
      
However, it must be stressed that the situation in de Sitter is very different from the black hole case since there exists no sensible notion of a ``stabilized" de Sitter. 
Instead, as discussed in~\cite{Dvali:2018ytn, Dvali:2021bsy}, the memory burden gives a consistency upper bound on the duration of classical de Sitter in terms of $t_{\rm M}$. 
Originally, the bound on the duration of the de Sitter state was derived in~\cite{Dvali:2013eja, Dvali:2014gua, Dvali:2017eba} from the self-entanglement of the de Sitter state caused by the back reaction from Gibbons-Hawking radiation. \\

   In this respect, we must notice that the memory burden effect completes a bigger picture of previously suggested mechanisms leading to a breakdown of the semi-classical description for a macroscopic system after a certain critical time, so-called {\it quantum break-time}~\cite{Dvali:2013vxa}. \\
  
 In particular, it has been suggested~\cite{Dvali:2012rt, Dvali:2013eja} that a black hole experiences quantum breaking while still being macroscopic. This happens the latest after the loss of about half of the initial mass. 
  It has also been argued that at this point, the black hole acquires a significant quantum hair, which could potentially lead to its stabilization~\cite{Dvali:2012rt}.  \\
 
 The concept of quantum break-time has also been applied  
to de Sitter~\cite{Dvali:2013eja, 
Dvali:2014gua, Dvali:2017eba}.
 However, unlike a black hole, which can happily continue 
 the existence beyond this point, 
 the quantum break-time represents a consistency upper 
 bound on the duration of any classical de Sitter state.  
 In particular, in any consistent inflationary theory, the 
 inflation must end before the corresponding 
 quantum break-time elapses \footnote{The
 quantum inconsistency of an eternally inflating Universe also follows from its incompatibility with the $S$-matrix formulation of quantum gravity~\cite{Dvali:2020etd}.}.  \\
 
 The memory burden effect strengthens these earlier 
 quantum break-time proposals for black holes
 and for de Sitter,   
 as it provides an additional engine for quantum-breaking~\cite{Dvali:2021bsy}. 
   It also makes the microscopic 
 dynamics of quantum breaking very explicit.  \\
 
  In general, due to its universal nature, 
   the memory burden phenomenon strongly affects the  
 systems with high capacity of information storage, leading 
 to a number of physical consequences.  \\

The rest of this paper is organized as follows. Next section describes the memory burden effect in general systems of enhanced memory storage capacity, which is then specialized  to the case of solitons and black holes in Sec.~\ref{sec:memoryburdensoliton} and~\ref{sec:memoryburdenBHs} respectively. Sec.~\ref{sec:extremality} focuses on the analogies and differences between memory burden and classical extremality, while Sec.~\ref{sec:numerics} is dedicated to numerical results showing examples of dynamical stabilization of solitons by their memory. Finally, Sec.~\ref{sec:implications} contains remarks on the phenomenological consequences of our findings as well as our outlook. Visuals of our numerical simulations can be found at the following \href{https://youtu.be/boDpRXJnT5E}{URL}.
 

\section{Essence of Memory Burden} \label{sec:essenceofmemory}

The memory burden effect was introduced in~\cite{Dvali:2018xpy}, where it was proposed that the loaded information pattern tends to stabilize the system carrying this information. 
A more detailed analysis of prototype systems was performed in two follow-up papers,~\cite{Dvali:2018ytn} and~\cite{Dvali:2020wft}, with the purpose of applying the memory burden effect to cosmology and black holes, respectively.
The last work concluded that the slowdown of the black hole decay due to the effect is imminent, the latest, by its half-decay. 
  
In this chapter, following~\cite{Dvali:2018xpy, Dvali:2018ytn, Dvali:2020wft}, we shall explain the essence of the memory burden phenomenon and give some helpful classification of its regimes.   
In order to achieve this,  we first go through the main universal characteristics of the systems with the {\it enhanced capacity of information storage}. 
This concept refers to the energetic efficiency of quantum information storage and was introduced in~\cite{Dvali:2017ktv,  Dvali:2017nis, Dvali:2018vvx}  at the level of basic  Hamiltonians. The reader can find the summary of the setup in~\cite{Dvali:2018tqi,  Dvali:2018xoc, Dvali:2021bsy}.  
     
The main characteristics of effective Hamiltonians describing systems of enhanced information capacity can be determined by categorizing quantum degrees of freedom according to the tasks they perform. 
The degrees of freedom shall be introduced as quantum oscillators in number representations.
These shall later be identified with different modes of the quantum fields.  
    
\subsection{Memory modes}
The first category of the degrees of freedom is the {\it memory modes}.  We shall introduce them as quantum fields denoted by $\theta^j$ where the index $j=1,2,....,M$  labels their  ``flavor".
The total number of species is $M$. 
The corresponding creation-annihilation operators  $\hat{a}_j^{\dagger},\, \hat{a}_j$ can satisfy either fermionic or bosonic oscillator algebras.    
For definiteness, we shall focus on the bosonic one:
\begin{equation} \label{algebra}    
  [\hat{a}_i, \hat{a}_j^{\dagger} ] \, = \, \delta_{ij} \,,
  ~~\, [\hat{a}_i, \hat{a}_j ] =0\,.    
\end{equation}   

The role of the memory modes is to store information. The information is stored in the patterns of their occupation numbers. These patterns are represented via ket-vectors in the Fock space, 
\begin{equation} \label{algebra2}    
\ket{p} = \ket{n_1,n_2,...,n_M} \,, 
\end{equation} 
where numbers $n_j \equiv  \bra{p} \hat{n}_j \ket{p}$ represents the eigenvalues of the corresponding number operators, $\hat{n}_j \equiv \hat{a}_j^{\dagger} \hat{a}_j$. 

The Hilbert sub-space formed by the vectors $\ket{p}$ shall be referred to as the {\it memory space}.  The dimensionality of it is defined by the range of the occupation numbers of the memory modes $n_j$. 
These can be subject to constraints.

An example of a simple constraint is to limit each occupation number to two possible values, $n_j =0, 1$.  
In this case, the free part of the memory system is described by the Hamiltonian of $M$ independent qubits, 
\begin{equation} \label{Hfree} 
\hat{H}_{\rm free}  \, = \, \sum_{j=1}^{M}\, m_j \, \hat{n}_j
\end{equation}
where $m_j$ are the energy gaps of the memory modes.  
The corresponding dimensionality of the memory space is $n_{\rm st} \, =\, 2^{M}$.   
In general, the dimensionality of memory space grows exponentially with the number of memory species $M$. Thus, the existence of a large number of memory species is an essential condition for the efficiency of information storage. 
However, it is not sufficient. 

Even with large $M$, the Hamiltonian (\ref{Hfree}) does not necessarily represent a system of enhanced information capacity. 
This is because the energy span of the memory space can be very large if the gaps $m_j$ are high.  

Therefore,  the second important characteristic is the energy cost of an information pattern, 
\begin{equation} \label{EPV}
E_{p}  = \sum_{j=1}^{M} m_j\, n_j \,, 
\end{equation} 
as well as the gaps between different memory patterns, 
$\ket{p}$ and $\ket{p'}$,  
\begin{equation} \label{PP}
E_{p} - E_{p'} \,. 
\end{equation} 

In short, the efficiency of the information storage by a system is determined by the density of states: the number $n_{\rm st}$, of states $\ket{p}$, that can fit within a physically meaningful smallest energy gap $\Delta E$. 
Usually, this is set by a typical uncertainty in the system's energy, such as the level-width. 
Naturally, many flavors $M$ and smaller gaps $m_j$ achieve more efficiency.

The resulting information storage capacity is quantified by the microstate entropy of the system defined as, 
\begin{equation}
S \equiv \ln(n_{\rm st}) \,. 
\end{equation}  

In quantum field theory (QFT), the limit to memory capacity is set by saturons~\cite{Dvali:2020wqi}: the objects with microstate entropy that saturates the upper bounds (\ref{AlphaSbound}) and (\ref{AreaSbound}) imposed by the validity of a given  QFT description. 

\subsection{Master modes: assisted gaplessness} 

The second category of the degrees of freedom are so-called {\it master modes}. 
We shall denote them by $\phi_{\alpha}$, with creation-annihilation operators,  
$\hat{a}_{\phi_{\alpha}}, \hat{a}^{\dagger}_{\phi_{\alpha}}$, 
where $\alpha = 1,2, ...,$ is their flavor index. 

The role of the master modes is to {\it assist}  the memory modes in becoming gapless. 
As opposed to the memory modes, which must come in a large number of flavors $M$, the number of master mode species can be much less.  

To illustrate the mechanism of the assisted gaplessness, and the resulting effect of memory burden,  a single flavor of the master mode is sufficient. The corresponding number operator shall be denoted by  $\hat{n}_{\phi} \equiv \hat{a}^{\dagger}_{\phi} \hat{a}_{\phi} $. 

With the above conventions, the effect of assisted gaplessness~\cite{Dvali:2017ktv,  Dvali:2017nis, Dvali:2018vvx, Dvali:2018tqi} can be illustrated by using the following simple prototype Hamiltonian~\cite{Dvali:2018xpy, Dvali:2018ytn, Dvali:2020wft,Dvali:2021bsy}, 
\begin{eqnarray}       
\label{Hint} 
\hat{H} \, &=& \, \hat{H}_{\rm ms} \, + \,  \hat{H}_{\rm mem} \,, \\ \nonumber 
{\rm with}:~ 
\hat{H}_{\rm ms} &\equiv& m_{\phi}  \hat{n}_{\phi} \,, \\ \nonumber   
\hat{H}_{\rm mem} &\equiv &  \left (1 -  \frac{
\hat{n}_{\phi}}{N_{\phi}} \right )^q  \sum_j m_j  
\hat{n}_j \,,  
\end{eqnarray}
where, for physical clarity, we have split the Hamiltonian into the master $\hat{H}_{\rm ms}$ and memory $\hat{H}_{\rm mem}$ parts, respectively. 

The parameters $m_{\phi}$ and $m_j$ represent the {\it intrinsic} energy (or mass) gaps of the master and memory modes, i.e., the gaps around the Fock vacuum state $n_{\phi} = n_j =0$. 
The number $N_{\phi} \gg 1$ is a large number that sets the coupling between master and memory modes as $1/N_{\phi}$.  
Notice that in QFT systems, for a given $N_{\phi}$, $M$ is bounded as,  
\begin{equation} \label{Mbound} 
M \leqslant N_{\phi}\,.
\end{equation}
This bound can be understood as a manifestation of a general bound 
\begin{equation} \label{Cbound} 
({\rm coupling}) \times ({\rm number~of~species}) \leqslant 1\,,
\end{equation}
violation of which invalidates the QFT description~\cite{Dvali:2020wqi}. Thus, within the validity of QFT, the maximal number of memory patterns is achieved for  $M \simeq N_{\phi}$.

The parameter $q$, which we take as some positive even number, requires clarification. 
This quantity parameterizes the functional dependence 
of the effective energy gaps 
of the memory modes on $n_{\phi}$. 
Of course, in general, this function can be more complicated. 
However, near the point of the assisted gaplessness, it is well-described by a monomial with a power $q$.  
In this form, the  assisted gaplessness takes place for the critical occupation number of the master mode, 
\begin{equation} \label{nphiCrit} 
n_{\phi} = N_{\phi} \,. 
\end{equation}

Indeed, if we keep the master mode in the Fock vacuum state,  $n_{\phi} = 0$, and form a memory pattern via the excitations of the memory modes, $\ket{p} = \ket{n_1,...,n_M}$, the cost of energy is   
     \begin{equation} \label{Pvac} 
E_p =  \bra{p}\hat{H}_{\rm mem} \ket{p} \, = 
\, \sum_j m_jn_j \,.      
\end{equation} 
This can be extremely high if $m_j$-s are large.  

However, the system possesses  
another state (\ref{nphiCrit}),  with the exact same memory pattern  as $\ket{p}$, 
but with a critically excited memory mode.
We shall denote it by $\ket{\tilde{p}}$, where tilde indicates 
the difference in $n_{\phi}$.  

On this state, the contribution to the energy 
from the memory mode-dependent 
part of the Hamiltonian is zero: 
    \begin{equation} \label{Pvac2} 
E_{\rm mem} \, = \,  \bra{\tilde{p}} \hat{H}_{\rm mem} 
\ket{\tilde{p}} = 0 \,.      
\end{equation} 
This is because the effective gaps for the
memory modes are no longer given by $m_j$, but rather by 
the quantities, 
\begin{equation} \label{OmegaEff}
\omega_j \, \equiv \,   
\left (1 -  \frac{
n_{\phi}}{N_{\phi}} \right )^q m_j \,,
\end{equation}
which vanish on a state with $n_{\phi} \, = \, N_{\phi}$.     

Thus, for the information pattern 
it is energetically favorable that the master mode, 
instead of being in the vacuum $n_{\phi} =0$, 
is in the critical state  $n_{\phi} = N_{\phi}$.  
This is the essence of the mechanism of the {\it assisted gaplessness}~\cite{Dvali:2017ktv,  Dvali:2017nis, Dvali:2018vvx, Dvali:2018tqi} (the term was coined in~\cite{Dvali:2018tqi}).

However, the price to pay is the energy of the master mode: 
     \begin{equation} \label{Pvac3} 
E_{\rm ms} \, = \,  \bra{\tilde{p}} \hat{H}_{\rm ms} 
\ket{\tilde{p}} = m_{\phi} N_{\phi} \,.      
\end{equation} 
Due to this, for a given information pattern $p$, 
the energetically optimal state is determined by the balance 
between the two entries $E_{\rm mem}$ and $E_{\rm ms}$. 
This fixes 
the occupation number of the master mode, 
$n_{\phi}$, to a certain optimal value. 
The energy difference 
between this optimal state 
and master mode's  Fock vacuum ($n_{\phi} =0$), both evaluated for the same memory pattern $p$,
determines the energy efficiency of the information storage. 

It is useful to quantify this efficiency by defining 
the memory-efficiency coefficient as the ratio of the actual cost of an information pattern 
$E_{\rm mem}$ to its cost in the master mode Fock vacuum: 
    \begin{equation} \label{Epsilon}  
\epsilon \, \equiv  \, \frac{E_{\rm mem}}{E_p}\,.   
\end{equation}

Moving away from the above optimal state, requires climbing an energetic barrier.  This creates a resistance against 
abandoning the state of the enhanced memory capacity. 
This is the essence of the memory burden effect~\cite{Dvali:2018xpy, Dvali:2020wft}. 

\subsection{Memory burden effect} 

 We now wish to give certain universal characteristics of 
 the memory burden effect~\cite{Dvali:2018xpy, Dvali:2018ytn, 
 Dvali:2020wft}. 
 
  \subsubsection{Generalities} 
 
 In order to understand the energetic balance leading to the 
 memory burden effect, let us minimize the Hamiltonian  (\ref{Hint1}) 
with respect to $n_{\phi}$ in a state with a memory 
pattern  $\ket{p} = \ket{n_1,...,n_M}$. 
The intrinsic (vacuum) energy  cost  of the pattern, 
$E_p$, is given by 
(\ref{Pvac}).

 Notice that for minimization we can use Bogoliubov approximation in which we treat the operator $\hat{n}_{\phi}$ as a 
 $c$-number $n_{\phi}$.
  This is justified, since the occupation number around 
  the states of interest is macroscopic and 
  the $c$-number approximation of the operator works 
  up to corrections of $1/n_{\phi}$
  (for more detailed discussion of $c$-number method 
  see~\cite{Dvali:2018tqi}).   

 We can distinguish the two regimes depending on whether 
 $E_p$ is above or below the following critical value, 
     \begin{equation} \label{EPcrit}  
 E_*\, \equiv \, \frac{1}{q} m_{\phi}N_{\phi}  \,.    
\end{equation}
  For
      \begin{equation} \label{ElargerEstar}  
E_p \,  \geqslant \,  E_*\,,    
\end{equation}
 the minimum of the energy of the system is achieved 
 for    
      \begin{equation} \label{nphiA}  
n_{\phi} \, = \, N_{\phi} 
\left (1 \, - \,  \left ( \frac{E_*} 
{E_p} \right )^{\frac{1}{q-1}} \right ) \,,   
\end{equation}
 and it is equal to, 
       \begin{equation} \label{EnphiA}  
E \, = \, m_{\phi} N_{\phi} 
 \left (1 \, - \,  \frac{q-1}{q} \left ( \frac{E_*} 
{E_p} \right )^{\frac{1}{q-1}} \right ) \,.   
\end{equation}
 Taking into account (\ref{EPcrit}) and 
 (\ref{ElargerEstar}),  we can easily see that this energy is less than the 
 would-be vacuum energy cost of the pattern $E_p$. 
   That is, in this case, it is energetically favourable 
 to stabilize the system in the state with non-zero 
 occupation number of the master mode $n_{\phi}$. 
 
 On the other hand, for 
      \begin{equation} \label{EPcritL}  
E_p \leqslant E_*\,,    
\end{equation}
 the minimum energy state is, 
      \begin{equation} \label{nphizero}  
 n_{\phi} = 0\,,      
\end{equation}
with the energy, 
\begin{equation} \label{EEstar} 
E \, = \, E_p \,.   
\end{equation}
 Of course, at the critical value  $E=E_*$, the  two regimes 
 give the same minimal energy 
 \begin{equation} \label{Emerge} 
E \, = \, E_p \, = \frac{1}{q} m_{\phi}N_{\phi}  \,. 
\end{equation}

 That is, if the vacuum energy cost of the pattern is below 
 the critical value (\ref{EPcrit}), it is not worth energetically 
 to keep the system in $n_{\phi} > 0$ state.  
 Otherwise, the system is stabilized in a state 
 with $n_{\phi} > 0$ by the memory pattern. This is the key ingredient of the memory burden effect. \\  

Notice, the Hamiltonian (\ref{Hint}) conserves the number of the master mode.  Correspondingly, the states with arbitrary 
$n_{\phi}$ represent  its eigenstates and do not evolve in time. 
Therefore, in order to observe the stabilizing effect of the memory burden dynamically, we must add interactions 
that do not conserve $n_{\phi}$ and allow for 
quantum transitions of $\phi$-quanta into some 
    ``external" degrees of freedom $\hat{b}^{\dagger}, \hat{b}$, 
    with energy gaps $m_b$.  
    This can be achieved by the inclusion of the interaction terms 
    of the form,  
   \begin{equation} \label{phibmixing}  
(\hat{b}^{\dagger})^{\beta} (\hat{a}_{\phi})^{\alpha} \, +   \,{\rm h. c.} \,,   
 \end{equation}  
where $\alpha, \beta$ are integers. 

  The dynamical memory burden effect is clearly 
  illustrated already in the simplest case $\alpha=\beta=1$,
  with $m_{\phi} = m_b$,  which can be solved analytically~\cite{Dvali:2018xpy}. The Hamiltonian (\ref{Hint}) is supplemented by    
    \begin{equation} \label{Hextra} 
\hat{H}_{\rm int} \, = \,  
 \frac{\tilde{m} }{\sqrt{N_{\phi}}} \, \hat{b}^{\dagger} \hat{a}_\phi \, +  \,
\frac{\tilde{m}^* }{\sqrt{N_{\phi}}} \, \hat{a}^{\dagger}_\phi \hat{b} \,   
\, +  \, m_{\phi} \hat{n}_{b} \,,    
 \end{equation}
 where $\hat{n}_{b} \equiv \hat{b}^{\dagger}\hat{b}$ 
 is the $b$-mode number 	operator and 
 $\tilde{m}$ is a complex parameter of the dimensionality of
 energy. Notice that $1/\sqrt{N_{\phi}}$ in the mixing terms 
 makes the parametrization consistent with the normalization of the coupling of the master mode in (\ref{Hint}).
  
  The above system can be solved in two ways.  
 We can first  find the minimum of energy for 
 a fixed total occupation number  
 \begin{equation} \label{Nphib} 
 n_{\phi} \,  + \,  n_b \, = \, N_{\phi} \,, 
\end{equation}	  
 using the Bogoliubov approximation for 
 $a_{\phi}$ and $b$ modes.  Given the constraint  
 (\ref{Nphib}), without any loss of generality, 
 we can parameterize these modes replacing the operators 
 by the $c$-number functions, 
  \begin{equation} \label{abBogoliubov} 
 \hat{a}_{\phi} \, = \, 
 \cos(\theta)\, e^{i \alpha_{\phi}} \sqrt{N_{\phi}} \,, ~
 \hat{b} \, = \, 
 \sin(\theta)\,e^{i \alpha_{b}}\sqrt{N_{\phi}} \,, 
\end{equation}	  
where $\alpha_{\phi}$ and $\alpha_{b}$ are phases and 
$\theta$ is an angle that parameterizes the distribution of the occupation number among $a_{\phi}$ and $b$-modes. 
   Notice that the phases $\alpha_{\phi}$ and $\alpha_{b}$ drop out of the minimization procedure, since they 
   align with the phase of $\tilde{m}$  and 
   give the negative over-all sign of the mixing term. 
    
     The total effective Hamiltonian becomes a function 
 of a single  variable $\theta$, 
        \begin{equation} \label{Htheta} 
\hat{H}  \, = \,  \sin(\theta)^{2q}\, E_p \, - \, |\tilde{m}| \,\sin(2\theta) \,.    
 \end{equation}
 It is clear that the larger is $E_p$, the closer is the minimum 
 of the energy to $\theta =0$, which implies 
 $n_{\phi} =N_{\phi}$ and $n_b =0$. 
 
  For $qE_p/|\tilde{m}| \gg 1$, the system gets stabilized 
 in the state
  of enhanced memory capacity with the depleted number 
  of master modes, $\Delta n_{\phi} \, = \, N_{\phi} \, - \, 
 n_{\phi}$, given by,   
          \begin{equation} \label{theta1} 
\frac{\Delta n_{\phi}}{N_{\phi}} \, = \, \sin^2{\theta} \simeq 
\left ( \frac{|\tilde{m}|}{qE_p} \right )^{\frac{2}{2q-1}} \,.    
 \end{equation}
  Correspondingly, this number gets transferred to the $b$-mode.   

On the other hand, for $E_p \, = \, 0$, the minimum 
is achieved for 
$\theta = \pi/4$ implying that the average occupation numbers 
satisfy, 
  \begin{equation} \label{theta2}
  n_{\phi}\, = \, n_b \,.
 \end{equation}
 These results are fully confirmed by the explicit quantum evolution of the system.  
 
    Indeed, following~\cite{Dvali:2018xpy, Dvali:2020wft}, 
  let us time-evolve the system from  the initial 
 state $n_{\phi} = N_{\phi}$ and $n_b =0$. 
  
    For $E_p =0$, the  occupation numbers   evolve in time  as, 
   \begin{equation} \label{Nsin} 
\frac{n_{\phi}(t)}{N_{\phi}}\, = \,   \cos^2\left (\frac{|\tilde{m}|}{\sqrt{N_{\phi}}}
t \right ) \,, ~~\, 
\frac{n_b(t)}{N_{\phi}}\,= \,\sin^2\left (\frac{|\tilde{m}|}{\sqrt{N_{\phi}}}
t \right ) \,,
 \end{equation}
 which, for the averaged values, reproduce (\ref{theta2}).
  
For  $qE_p/|\tilde{m}| \gg 1$,  the story is very 
different.  The oscillation amplitude is now suppressed, 
so that the average value of $\Delta n_{\phi}$ is
given exactly by (\ref{theta1}).  
  For a detailed numerical analysis, see~\cite{Dvali:2020wft}. \\

 Of course, the above example of the system's decay represents an oversimplified 
 toy model, with the main purpose of illustrating the system with 
 a swift memory burden effect.  The time evolution must be taken with a 
 grain of salt, especially in the later oscillatory period. 
 
For QFT systems,  such as a black hole or a soliton, submerged in infinite space,  the master mode can decay into a continuum of the external $b$-modes. These correspond to different momentum (or spin) eigenstates of outgoing radiation. In such cases, the time-evolution of the system 
is dissipative rather than oscillatory.    Regardless, 
as long as the system maintains the information pattern with 
$E_p \gg m_{\phi}$, it is subjected to memory burden, latest for  
 $\Delta n_{\phi}/N_{\phi} \sim 1$. That is,
  the memory burden phase starts latest by the time the system gets rid of an order-one fraction 
 of its initial energy. 
 
  The way of avoiding the memory burden would be 
 for the system 
  to get rid of the information pattern very fast~\cite{Dvali:2018xpy}. However, 
  in systems of enhanced information capacity, 
  due to the assisted gaplessness, this is an extremely suppressed process~\cite{Dvali:2020wft}. 
  For a detailed numerical analysis
 of complex  prototype systems, demonstrating this outcome, 
the reader is referred to the above paper.  \\
   
  We now move to discussions of different  parameter regimes and distinguish the two  extreme realizations of the memory burden effect.     

  \subsubsection{Type-$I$ regime} 
  
   The first regime, which we call Type-$I$, takes place when the intrinsic frequencies of the  master and memory modes are of the same order.   
    We shall take them to be given by  an universal mass gap $m$: 
    \begin{equation} \label{type1}
  m = m_{\phi} = m_j\,.  
  \end{equation}
  The Hamiltonian (\ref{Hint}) becomes,  
    \begin{equation} \label{Hint1} 
 \hat{H} \, = \, m \, \hat{n}_{\phi} \, + \, m \, 
\left (1 \, - \,  \frac{
 \hat{n}_{\phi}}{N_{\phi}} \right )^q \sum_{j=1}^{M}  
  \hat{n}_j \,.  
\end{equation}
 Notice that it is invariant under an arbitrary 
 $U(M)$ transformation of the memory 
 modes: $\hat{a}_j \rightarrow U_{jk}\hat{a}_k$. 

 Let us evaluate the Hamiltonian on a memory pattern, 
 with total occupation number of all species given by, 
 \begin{equation} \label{NmemG}
 \sum_{j=1}^{M} n_j \,  =\, N_{\rm G}\,.
 \end{equation}  
 Due to 
 $U(M)$ symmetry this number can be arbitrarily 
 redistributed among $M$ memory modes, which creates 
 the following number of degenerate 
 microstates~\cite{Dvali:2020wqi},
 \begin{equation}
\label{eq:numberofstatesg}
   n_{\rm states}\simeq \left(1 +  \frac{N_{\rm G}}{M} \right)^{M} \left(1 +  \frac{M}{N_{\rm G}} \right)^{N_{\rm G}} \,.
\end{equation}
The above expression represents a binomial coefficient $\binom{N_{\rm G}}{M}$ evaluated using Stirling approximation for large $N_{\rm G}$ and 
$M$.

 This degeneracy can be understood in terms of the Goldstone 
 phenomenon, since the occupation number of modes 
 $N_{\rm G}$ breaks the $U(M)$ symmetry spontaneously down 
 to $U(M-1)$.  
 
  The memory pattern back reacts on the master mode. 
  In order to understand this back reaction, 
  following~\cite{Dvali:2018vvx},
we minimize the Hamiltonian  (\ref{Hint1}) 
with respect to $n_{\phi}$. This gives, 
    \begin{equation} \label{minN}  
\left (1 \, - \,  \frac{
 n_{\phi}}{N_{\phi}} \right )^{q-1} = \frac{N_{\phi}}{qN_{\rm G}} \,.   
\end{equation}
 We see that for $N_{\rm G} \geqslant N_{\phi}$, 
 $n_{\phi} \simeq N_{\phi}$.
  Plugging (\ref{minN}) in the generic 
 expression (\ref{OmegaEff}) for the effective gaps of the memory modes, we get,   
   \begin{equation} \label{Om1}  
\omega \,
= \,
 m  \left ( \frac{N_{\phi}}{qN_{\rm G}} \right )^{\frac{q}{q-1}} \,.   
\end{equation}
 In particular, for $q \gg 1$, 
 we obtain, 
    \begin{equation} \label{Om2}  
\frac{\omega}{m} \, \simeq \, 
  \frac{N_{\phi}}{qN_{\rm G}}  \,.   
\end{equation}
Notice also that the contribution to the energy 
from the master and memory modes 
    \begin{equation} \label{EmemEms}  
E_{\rm ms} \, = \, m\, N_{\phi}\,,~~ 
E_{\rm mem} \, = \,  \omega \, N_{\rm G}  \,,   
\end{equation}
 relate as 
      \begin{equation} \label{EEq}  
E_{\rm mem} \, \simeq  \, \frac{1}{q} E_{\rm ms} \,.   
\end{equation}
  From (\ref{EmemEms})  and (\ref{Om2}), 
 it is clear that 
 in type-$I$, the memory-efficiency coefficient is: 
        \begin{equation} \label{Eps1}  
\epsilon_{I} \, \simeq  \, \frac{\omega}{m} \simeq \,
\frac{N_{\phi}}{qN_{\rm G}} \,.   
\end{equation}

 In realistic systems, due to interactions with external
 degrees of freedom, the occupation number $n_{\phi}$
 can change in time. In the previous section 
 we reproduced an example of~\cite{Dvali:2018vvx}
 in which the transition takes place into a single $b$-mode. In this case, the behaviour 
 is oscillatory. However, when the number of external decay channels is large, 
 the inverse transitions practically never happen and 
 $n_{\phi}$ decreases irreversibly.  
 
 For example, this is what happens
 in black holes~\cite{Dvali:2020wft}, where the master mode 
 gets depleted into outgoing Hawking particles~\cite{Dvali:2011aa}.  
In such cases,  the system dynamically 
evolves until it gets stabilized (or semi-stabilized)  
 by the memory burden effect.  
 
   In other words, the information drives the system 
   towards the state in which the energy cost of the information pattern 
   is optimal. 
   Correspondingly, the system resists against the attempts of 
  driving it away from such a state.     
 We shall later discuss this dynamics in solitons and in black holes and 
 compare them.  

  \subsubsection{Type-$II$ regime} 
   We shall now consider the regime 
   in which the intrinsic energy gaps of the memory modes 
   are  much higher than that of the master mode: 
  \begin{equation} \label{type2}
  m_j \gg m_{\phi}\,.  
   \end{equation}
   We shall refer to it as  type-$II$.
   In this case the minimization of the Hamiltonian 
    with respect to $n_{\phi}$ gives, 
       \begin{equation} \label{minN2}  
\left (1 \, - \,  \frac{
 n_{\phi}}{N_{\phi}} \right )^{q-1} = m_{\phi}\frac{N_{\phi}}{qE_p} \,.   
\end{equation}
   Taking into account 
  (\ref{Cbound}) and (\ref{type2}), we notice that 
  for a memory pattern with $N_{\rm G} \gtrsim M$, 
  the r.h.s., of the above equation is much less than one. 
  Thus,  similarly to type-$I$ case,  we have 
  $n_{\phi} \simeq N_{\phi}$. 
  Correspondingly, the energy cost of the master mode is 
        \begin{equation} \label{Ems2}  
  E_{\rm ms} \, \simeq \, m_{\phi} N_{\phi} \,. 
 \end{equation} 
  At the same time, the actual  energy cost of the memory pattern is, 
        \begin{equation} \label{Emem2}  
 E_{\rm mem} = \left (m_{\phi}\frac{N_{\phi}}{qE_p} \right )^{\frac{q}{q-1}}
 E_p
 \simeq \frac{1}{q} m_{\phi} N_{\phi}  \,.     
\end{equation}
 Thus,  we arrive to the same 
 relation (\ref{EEq}) as in type-$I$ case. 
  
   However, the efficiency coefficient $\epsilon$ is now 
   extra-suppressed.  For example, taking all 
   $m_j$ to be set by a single scale, $m_j = m \gg m_{\phi}$, 
   we get 
           \begin{equation} \label{Eps2}  
\epsilon_{II} \, \equiv  \, \frac{N_{\phi}}{qN_{\rm G}} \frac{m_{\phi}}{m} \,.    
\end{equation}
 This is suppressed with respect to type-$I$ case 
 (\ref{Eps1}) by an additional factor $m_{\phi}/m$. 
 
 The basic lesson is the following:  \\
 
 {\it  Increasing the vacuum 
 gaps of the memory modes with respect to the master mode, 
 the system becomes more and more efficient 
 in information cost. Correspondingly, the memory burden effect 
 is stronger. } 
       
\section{Memory Burden in Solitonic Saturon} \label{sec:memoryburdensoliton}

   We shall now study the memory burden 
   effect in solitonic saturons.
     Following~\cite{Dvali:2020wqi}, we first introduce 
    an example of a saturon in form of a vacuum bubble.   
 Next, we discuss their stabilization via memory burden 
 effect closely following~\cite{Dvali:2021tez}. 

The bubble is a solution 
in a renormalizable  $SU(N)$-invariant theory   
of a scalar field $\Phi$ in the adjoint representation 
with the following Lagrangian density, 
\begin{equation}
\label{eq:lagphi}
    \mathcal{L}[\Phi]= \frac{1}{2}{\rm{Tr}}\,(\partial_\mu \Phi)(\partial^\mu \Phi) - 
   V(\Phi) \,, 
 \end{equation}
  where the potential is chosen as, 
 \begin{equation}     
  V(\Phi) \, = \,  \frac{\alpha}{2}{\rm{Tr}}\left(f\Phi -\Phi^2 +\frac{I}{N}{\rm{Tr}}\,\Phi^2\right)^2 \,.
\end{equation}
Here $\alpha$ is a coupling constant and $f$ is a scale.
 $I$ is an unit matrix in $SU(N)$-space. 
 
  The validity of QFT description imposes the following bound 
  on the parameters of the model, 
  \begin{equation} \label{Ubound} 
   \alpha N \lesssim 1 \,.
   \end{equation}
   Basically, the quantity $\alpha N$, which is analogous to 't Hooft coupling, must not exceed the critical value.  
   In the opposite case ($ \alpha N \gg 1$),  $\Phi$  no longer represents a good QFT degree of freedom. This is
   unambiguously signalled by the breakdown of the loop-expansion, as well as, by the saturation 
   of unitarity in multi-particle scattering amplitudes~\cite{Dvali:2020wqi}.     
    Notice that after we map the above theory on a generic Hamiltonian  of enhanced memory capacity (\ref{Hint}), it will  become clear that the bound  (\ref{Ubound})
    represents a particular case of 
   the bound (\ref{Cbound}).   
  
The system~\eqref{eq:lagphi} has multiple degenerate vacua satisfying the condition
\begin{equation}
    f\, \Phi_a^b-\left(\Phi^2\right)^b_a + \frac{\delta^b_a}{N}\,{\rm{Tr}}\,\Phi^2\,=\, 0 \,.
\end{equation}
We shall focus on a pair of neighbouring vacua. 

 In the first one, the  vacuum  expectation value (VEV) is 
 $\Phi =0$ and  the global  $SU(N)$-symmetry is unbroken.  
This vacuum exhibits a mass gap 
\begin{equation} \label{MassGap} 
m = \sqrt{\alpha}f \,,   
\end{equation} 
 which sets the minimal energy cost for 
all particle excitations. 

 In the second vacuum of our interest, the  VEV is, 
\begin{equation}
\label{eq:ans}
    \Phi_0 \, = \, f \frac{1}{\sqrt{N(N-1)}}{\rm{diag}}\left(N-1,-1,...,-1 \right) \,,
\end{equation}
and $SU(N)$-symmetry is spontaneously broken 
down to $SU(N-1)\times U(1)$.
Due to spontaneous breaking of symmetry,   in this vacuum we have
the set of massless Goldstone bosons     
\begin{equation}
\theta^j(x) \,, \quad j=1,2, \dotsc , M \,, 
\end{equation}
corresponding to broken generators $T^{j}$. 
Their total number is 
 \begin{equation}
M \, = \, 2(N-1) \,. 
\end{equation}
 These Goldstones form a  
 fundamental representation of the $SU(N-1)$ group of complex dimensionality $N-1$ with non-zero charge under $U(1)$. 
 
 The effective Lagrangian of the Goldstone bosons has the following form, 
  \begin{equation}
\mathcal{L}_{\text{eff}} \, = \, 
 f^2\left(\partial_{\mu} {\mathcal U}(x) \right) \left(\partial^{\mu}  {\mathcal U}(x)  \right)  \, ,
\end{equation}
with 
\begin{equation} 
\label{MUU}
  {\mathcal U}(x)  \equiv  f^{-1} U(x)^{\dagger}\Phi_0 U(x)\,,
  \end{equation}
  where $U(x)$ is the space-time dependent
  $SU(N)$-transformation matrix, 
 \begin{equation}
\quad U(x) = \exp \left[ - i \theta^j(x) T^j \right] \,.
\end{equation}
  Up to second order in $\theta^j(x)$-s
  and leading order in $1/N$, the Goldstone 
  Lagrangian can be written as, 
   \begin{equation}
\mathcal{L}_{\text{eff}} \simeq \frac{1}{4} f^2 \sum_j\left(\partial_{\mu} \theta^j \right) \left(\partial^{\mu} \theta^j \right) \, .
\end{equation}

\subsection{Vacuum Bubbles} 
    
    Due to the degeneracy of vacua, there also exist the domain wall configurations separating 
 them,  
 \begin{equation}
\label{BubbleN}
    \Phi (x)  = \phi(x) \frac{1}{\sqrt{N(N-1)}}{\rm{diag}}\left(N-1,-1,...,-1 \right)\,,
\end{equation}
  where the function 
  $\phi(x)$ interpolates between  $0$ and $f$. 
  We shall be interested in a spherically symmetric 
  bubble (closed wall) of $\phi=f$ vacuum 
  embedded in $\phi=0$ one. 
  
  Let us first consider the following configuration, 
   \begin{equation}
\label{oscillon}
    \Phi  = \frac{\phi(r,t)}{f} \Phi_0 \,.
\end{equation}
  The field $\phi(r,t)$ satisfies the following 
 equation of motion, 
 \begin{equation} \label{EQR}
\partial_t^2 \phi  -\partial_r^2\phi  -\frac{2}{r} \partial_r \phi 
+ \frac{\partial V}{\partial \phi} \, = \, 0 \,,
\end{equation}
where, 
 \begin{equation} \label{VPHI}
 V(\phi) \, = \, \dfrac{\alpha}{2} \phi^2 \left(\phi - f \right)^2 \, .
\end{equation}
   The boundary conditions  are such   
  that for  $t=0$,   $\phi(r=\infty) =  0$ and 
  $\partial_t\phi |_{t=0} = \partial_r\phi |_{r=0} =0$. 

If the initial radius of the bubble satisfies
 $R \gg m^{-1}$, this configuration
can be approximated by 
\begin{equation} \label{LargeB} 
\phi(r)= \dfrac{f}{2} \left[ 1 + \tanh \left( \dfrac{ m(R -r)}{2} \right) \right] \, .
\end{equation}
  
  The bubble wall has a thickness $\delta_w \simeq m^{-1}$ 
  and a tension, $\sigma \simeq m^3/(6\alpha)$.   
  Order of magnitude wise, this remains true also 
  for thick-walls bubbles for which $R \sim \delta_w$.  
    
In the interior of the  bubble, the $SU(N)$ symmetry is broken spontaneously down to $SU(N-1)\times U(1)$.  
 Due to this, there are   
$M = 2(N-1)$ gapless 
Goldstone species localized within the bubble. 
These species do not exist outside of the bubble, since 
$SU(N)$-symmetry is restored there. 

As a consequence, the bubble has a large
microstate degeneracy. 
 This degeneracy follows from spontaneously broken 
 $SU(N)$-symmetry, since a bubble obtained from 
 (\ref{BubbleN}) by an arbitrary $SU(N)$-transformation
  $U$, 
\begin{equation}
\label{eq:fullans}
    \Phi(x) \rightarrow U^{\dagger}\, \Phi(x) \, U,
\end{equation}
represents a classical solution of the same energy. 

  In quantum theory, the number of degenerate microstates is not 
 infinite, since only the orthogonal states must be counted. 
  These states are obtained by redistribution of the constituent 
  quanta of the bubble among different  $SU(N)$-flavors  
  which (in large-$N$) amounts to the following 
number of degenerate microstates~\cite{Dvali:2020wqi}, 
\begin{equation}
\label{eq:numberofstates}
   n_{\rm states}\simeq \left(1 +  \frac{s(R)}{2N} \right)^{2N} \left(1 +  \frac{2N}{s(R)} \right)^{s(R)} ,
\end{equation}
 where the quantity $s(R) = 4\pi (R\,m)^3/\alpha$ 
 is the time-averaged space integral of $\phi^2(r)$.
 This quantity effectively measures the mean occupation number of constituent quanta of the bubble soliton  
 viewed as a coherent state~\cite{Dvali:2015jxa}. 
 
     Taking the maximal value of $N$ given by
    (\ref{Ubound}), for  thick-wall bubbles $R \sim 1/m$, 
     we get the following expression  
    for entropy, 
  \begin{equation}
 \label{SBubbleN}
   S \sim N  \sim \frac{1}{\alpha} \,. 
\end{equation} 
It is clear that this expression represents the area
($\sim 1/m^2$)  in units of the Goldstone scale $f$.  
 Thus, for thick wall bubbles the entropy saturates the area-law 
 bound~(\ref{AreaSbound}). 

 For $R \gg 1/m$,  the entropy is, 
 \begin{equation}
 \label{SlargeB}
   S\,  \sim \, N \ln(mR) \,. 
\end{equation}

  Due to the above degeneracy, the  bubble represents a system of enhanced capacity of the memory 
storage. The memory modes are Goldstone modes
$\theta^j$. The master mode is the radial mode, $\phi(x)$.
 
  Setting other components to zero, 
 the effective Lagrangian of Goldstone modes 
and of the radial mode $\phi$ is given by 
  \begin{equation}
\mathcal{L}_{\text{eff}} \, = \, \frac{1}{2} \partial_{\mu} \phi 
\partial^{\mu}\phi  + \frac{1}{2} \phi^2\left(\partial_{\mu} {\mathcal U}(x) \right) \left(\partial^{\mu}  {\mathcal U}(x)  \right) 
- V(\phi) \,.
\end{equation}
 The Goldstone modes are well-defined exceptionally 
 for $\phi \neq 0$. 

\subsection{Stabilization by memory burden} \label{subsection:stabilizationbymemory}

We can now observe the stabilization by the memory burden effect. In this, we shall closely follow~\cite{Dvali:2021tez}, where this effect was studied.    
  This effect requires that a non-zero information pattern 
  is stored in non-zero frequency excitations of Goldstone modes.   That is, some of the 
  Goldstone modes
 (i.e., the memory modes) of non-zero frequencies are occupied to numbers $n_a$. 
  \\
 
On such a state, $\ket{p} \equiv \ket{n_1,n_2,...,n_M}$, the measure of the memory burden effect is the expectation value, 
   \begin{equation}
\bra{p} \left(\partial_{\mu} {\mathcal U}(x) \right) \left(\partial^{\mu}  {\mathcal U}(x)  \right) \ket{p} \,.
\end{equation}

In order to understand the effect, let us first consider 
a classical configuration (\ref{MUU}) with,
   \begin{equation} \label{UUU}  
     U = e^{i t\omega_j T^j}\,. 
  \end{equation}    
    It is easy to see that this gives the equation, 
 \begin{equation} \label{EQR2}
\partial_t^2 \phi \, - \, \partial_r^2\phi  -\frac{2}{r} \partial_r \phi 
-  \omega^2 \phi +  \frac{\partial V}{\partial \phi} \, = \, 0 \,,
\end{equation}
 with 
    \begin{equation} \label{Omegas}  
     \omega^2 \, \equiv \,  \sum_{j=1}^M \omega_j^2 \,. 
  \end{equation}  
 Different patterns $(\omega_1, \omega_2, ...)$  can be obtained from one another by $SU(N-1)$ transformation, which leaves the  $\Phi_0$ invariant.  

 The above equation always has a stationary bubble solution with time-independent $\phi(r)$. The asymptotic values are 
 $\phi(\infty) = 0$ and $\phi(0) \neq 0$. 
  Since various patterns $(\omega_1, \omega_2, ..., \omega_M)$ are related by a symmetry transformation, it is sufficient to 
  discuss the case $\omega^j = \delta^{j1}\omega$
  and later generalize to different patterns.  
  
   In the thin-wall limit,  $R \gg 1/m$, the profile of the bubble is given by (\ref{LargeB}), where
 the radius $R$ can be determined by extremizing the energy of 
 the bubble as function of $R$, 
 \begin{equation} \label{TWEnergy}
E=\dfrac{2 \pi}{3 \alpha}m^3 R^2 (1-\dot{R}^2)^{-1/2} + \dfrac{2\pi}{3 \alpha} m^2 \omega^2 R^3 \, ,
\end{equation}
 subject to the condition that the quantity 
 \begin{equation} \label{Charg}
Q= \dfrac{2\pi }{3} f^2 \omega R^3= \dfrac{2\pi }{3\alpha} m^2 \omega R^3 \,, 
\end{equation}
represents a conserved charge. This determines the 
radius of the stationary bubble as,    
\begin{equation} \label{thinwallR}
  R_0 = \dfrac{2}{3}\frac{m}{\omega^2} \,.
  \end{equation} 
 Correspondingly, the energy of the bubble is, 
    \begin{equation} \label{thinwallE}
 E_{0} 
  = \dfrac{\omega}{\alpha}\,  \frac{m^5}{\omega^5} \, \left (\frac{40\pi}{81} \right ) \, .
 \end{equation} 

 As already discussed in~\cite{Dvali:2021tez}, the stationary bubble solution
can be mapped on a $U(1)$ non-topological soliton 
or a $Q$-ball~\cite{Lee:1974ma,Friedberg:1976me, Coleman:1985ki}
formed by a complex scalar with modulus $\phi(r)$ 
and the $U(1)$-charge $Q$ given by~\eqref{Charg} (for some implications of $Q$-balls, see, e.g.,~\cite{Safian:1987pr,Lee:1991ax,Kusenko:1997ad,Dvali:1997qv,Kusenko:1997si, Kusenko:1997vi, Kim:1992mm, Volkov:2002aj}).  
  However, despite the similarity of classical solutions, 
such a soliton will not exhibit a significant microstate degeneracy
and will not represent a system of enhanced capacity of information storage. Correspondingly, the charge $Q$ cannot be 
associated with any information pattern.
  For a significant microstate degeneracy, having $SU(N)$-symmetry is crucial. Only in such a case the total charge 
  $Q$ can be distributed among an exponentially large number 
  of memory patterns. 

 The stationary bubbles carrying arbitrary patterns are obtained 
by $SU(N-1)$ transformations and have identical macroscopic 
features such as the energy and the radius.   However, they have different characteristics 
with respect to an asymptotic probe with fixed $SU(N-1)$ quantum numbers relative 
to which the rotation is performed. 
 
In other words, the information is stored in $SU(N-1)$-rotations of a pattern 
 relative to a fixed observer.  Obviously, all rotated bubbles have identical energies 
 but they are distinguished by the observer who can perform a scattering experiment 
 using a fixed probe.  
 
The stationary bubble represents a bubble stabilized by 
the memory burden effect. In terms 
of occupation numbers of Goldstones, $n_j$, 
the quantities  $\omega_j$ read, 
\begin{equation} 
\omega_j^2 \, = \, \omega \, n_j \,\frac{3}{2\pi f^2R^3_0} \,.
\end{equation} 
From the quantum perspective, it is illuminating  to express  the  energy of the bubble in terms 
   of the occupation numbers of the radial (master) mode, 
   $N_{\phi}$, and  of the Goldstone 
   modes, 
   \begin{equation} \label{NG}
   N_{\rm G} \equiv \sum_{j=1}^{M} n_j \,. 
\end{equation} 
  Splitting the energy of the stationary bubble (\ref{thinwallE}) into two contributions, we have: 
      \begin{equation} \label{Esplit}
 E_{0} \, = \, E_{\text{mem}} \, + \, E_{\text{ms}} \, ,
 \end{equation} 
 where  
 \begin{equation} 
 E_{\text{mem}}  = \omega N_{\rm G} \, , ~~ \text{with}, \, ~~ N_{\rm G} \equiv \dfrac{1}{\alpha} \dfrac{m^5}{\omega^5} \left(\dfrac{16\pi}{81}\right ) \, ,
\label{Ememory}
\end{equation}
is the energy of memory (Goldstone) modes, whereas  
  \begin{equation} 
  \label{Emaster} 
 E_{\text{ms}} =  m  N_{\phi} \, , ~~ \text{with}, \, ~~ N_{\phi} \equiv \dfrac{1}{\alpha} \, \dfrac{m^4}{\omega^4} \,
 \left(  \dfrac{8\pi}{27} \right ) \, , 
\end{equation}
is the energy of the master (radial) mode $\phi(r)$.  
At the stationary point we have  
    \begin{equation} \label{EC}
 E_{\text{mem}} = \dfrac{2}{3} E_{\text{ms}} \, . 
\end{equation}
Notice that this reproduces the generic equation
(\ref{EEq}) with $q=3/2$. 

  From (\ref{Charg}) and (\ref{Ememory}) it is easy to notice that we 
  have, 
\begin{equation}\label{QandNG}
   Q = N_{\rm G} \,.
 \end{equation}   
That is, the classical charge in quantum theory counts the 
total occupation number of Goldstone modes. 
  Various memory patterns are obtained by distribution of 
  $N_{\rm G}$ among $M = 2(N-1)$ flavors of Goldstones. 

Next, notice that the memory pattern with the same 
$N_{\rm G}$ in the 
$SU(N)$-invariant vacuum 
would cost 
\begin{equation} \label{EpNVac}
E_p = mN_{\rm G} \,, 
\end{equation} 
since in this vacuum the minimal energy gap of any elementary 
excitation with $SU(N)$ quantum number is $m$.  

 Taking this into account,  we can evaluate the 
memory-efficiency coefficient (\ref{Epsilon}), which gives, 
        \begin{equation} \label{EpsB}  
\epsilon \, =  \, \frac{\omega}{m} \simeq \,
\frac{2N_{\phi}}{3N_{\rm G}}  \,.  
\end{equation}
In the last equality we have expressed $\omega/m$ 
through occupation numbers via (\ref{Ememory}) and 
(\ref{Emaster}).  
This reproduces the equation (\ref{Eps1})
 with  $q=3/2$. We thus see that the vacuum bubble 
 realizes the type-$I$ memory burden effect, 
 with $q=3/2$.  
 
 The type-$I$ nature of the memory burden 
 is not surprising, since 
 in the $SU(N)$ invariant vacuum, the energy gaps 
 of all excitations are equal to $m$. 
 
  The stationary bubble represents a bubble stabilized by the 
  memory burden. Notice that for such a bubble 
  we have, 
      \begin{equation} \label{QNG} 
Q = N_{\rm G}  =  \dfrac{\pi}{\alpha} (mR_0)^{5/2}
 \left(\dfrac{2}{3}\right )^{3/2} \, .
\end{equation}
  A bubble with an insufficient memory burden will not be stationary. In particular, such is a bubble 
  with the radius  larger than the critical value $R \gg R_0$. 
  Obviously, for such a bubble, 
     \begin{equation} \label{NGV} 
N_{\rm G} \ll \dfrac{\pi}{\alpha} (mR_0)^{5/2}
 \left(\dfrac{2}{3}\right )^{3/2} \, ,
\end{equation}
and 
    \begin{equation} \label{ECV}
 E_{\text{mem}} \ll  \dfrac{2}{3} E_{\text{ms}} \, . 
\end{equation}
The memory burden is insufficient for stabilizing such a bubble. 
Instead,  as shown in~\cite{Dvali:2021tez},
 the bubble starts to collapse and oscillate emitting 
energy.  This decreases the occupation number of the master 
mode.  This process shall continue until the 
 balance is restored to (\ref{EC}). At this point the 
 bubble gets stabilized by the memory burden effect.   
 
  In particular,  the initial bubbles of equal energies but 
  different values of $N_{\rm G}$ will evolve into the stabilized bubbles of different energies as we shall discuss next. 

  \subsection{The energy splitting by memory burden}\label{subsec:energysplitting}

 We wish to discuss the following important effect
 which is characteristic of the phenomenon of memory burden stabilization. 
 Namely, the initially-degenerate energy states with different 
memory patterns $N_{\rm G} \neq N_{\rm G}'$, upon stabilization by the memory burden, split in energies. 

 Let us start with two initial bubble states with 
different values of the Goldstone occupation numbers, 
$N_{\rm G}$ and $N_{\rm G}'$ respectively. 
We assume that both numbers are subcritical 
(\ref{NGV}). Thus, initially the memory burdens are dynamically insignificant. According to (\ref{Ememory1}), the contribution  to the energy from the memory modes is, 
   \begin{equation} 
 E_{\text{mem}}  = \omega N_{\rm G} \, \sim 
 N_{\rm G} \sqrt{\frac{m}{R}}\,. 
\label{Ememory1}
\end{equation}
 Notice that order-of-magnitude wise this expression holds 
 also for  (\ref{NGV}), since $\omega \sim  \sqrt{\frac{m}{R}}$. 

 Correspondingly, the energy splitting between the two bubble states with fixed difference, $\Delta N_{\rm G} \equiv  N_{\rm G} - N_{\rm G}'$,
    \begin{equation} 
 \Delta E  \,\sim \, 
   \Delta N_{\rm G} \,\sqrt{\frac{m}{R}} \, ,
\label{DEmemory}
\end{equation}
vanishes for $R\rightarrow \infty$. 
 Thus, for sufficiently large and equal radii, the two bubble states are nearly degenerate.  

 Notice that the states with different $N_{\rm G}$ are fully legitimate microstates of the same macrostate.
   Semi-classically, the detection of the
 Goldstone charge is not possible. In quantum theory
 any attempted measurement is suppressed by powers of 
 $\Delta N_{\rm G}/N_{\phi}$, which vanishes in the semi-classical limit. 
 Thus, the microstates with different $N_{\rm G}$ are equally indistinguishable semi-classically as the microstates 
 with equal $N_{\rm G}$, but relatively rotated in $SU(N)$-space.

Now, since the occupation numbers of Goldstones, 
  $N_{\rm G}, N_{\rm G}'$, are subcritical 
 (\ref{NGV}), the bubbles will evolve in time  by oscillating and emitting  energy until they get stabilized by the memory burden effects. 
  Since during this process the radii  diminish and correspondingly the Goldstone frequencies ($\omega$) grow,   
  so does the energy splitting between the two bubble states.     

At the stabilization point, the bubble energies can be highly split
if the relative difference of Goldstone numbers is large. 
For example, for  $\Delta N_{\rm G}/N_{\rm G} \sim 1 $, the energy splitting between the two bubble ``remnants" is of order their masses. Of course, the energy of a remnant plus radiation is the same in both cases. \\

To summarise: For a soliton that has not yet entered 
the memory burden stabilization phase (i.e., an under-burdened soliton with (\ref{NGV})), we can distinguish the two types of degenerate microstates:  1) the microstates that have the same memory burden 
$N_{\rm G}$; and 2) the microstates with different values of $N_{\rm G}$. 

The first category belongs to the same orbit in 
$SU(N)$-space and, therefore, remains degenerate throughout the 
bubble evolution. However, the energies of the second category
undergo the splittings according to (\ref{DEmemory}). 
\\
 
\begin{figure}
 \centering
    \includegraphics[width=.4\textwidth]{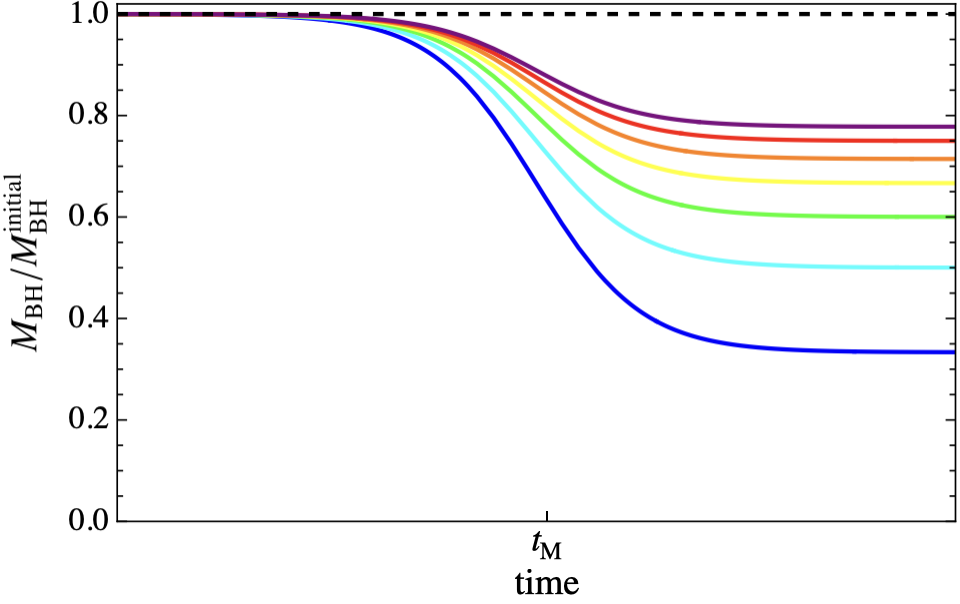}
    \caption{Schematic view of the splitting of energies of 
  the microstates of either a soliton or a black hole,  time-evolved from initially (nearly) degenerate microstates with different memory patterns. The solid lines represent the energies  of the remnant states, whereas the black dashed line represents the total energy of the system 
    (the remnant mass plus the energy of emitted radiation). Notice that each solid line is not a single state but accounts for all the states that remain degenerate due to identical memory burdens.  
 The degeneracy can be exact due to symmetry. For example, 
 such are the vacuum bubble microstates related by $SU(N)$-symmetry. 
 The line can also 
 exhibit a finer-structure, in case the symmetry is only approximate.
    }
    \label{fig:schematic_memory}
\end{figure}

The above situation is generic for the memory burden
  effect.  In particular, as we shall discuss, the same phenomenon takes place 
in black holes.  There too, after entering the memory burden 
domination  phase, the masses of initially-degenerate black 
holes with different information patterns can become vastly different - see Fig.~\ref{fig:schematic_memory} for a schematic visual. 
We shall also demonstrate this effect numerically in Sec.~\ref{sec:numerics}.

Finally, we wish to stress that while the memory burden stabilizes 
 the systems, the converse is not excluded: 
 the stable systems with very little information 
 capacity can happily exist. 
 An example of such a system would be the $Q$-ball 
 with the $U(1)$-charge. Although the $U(1)$-charge carries very little information, it nevertheless can stabilize the system dynamically.  
 
 The important thing about the systems with high information content (such as saturons) is that they have no choice: the memory burden effect stabilizes them necessarily. 

  \section{Memory burden in black holes} \label{sec:memoryburdenBHs}
 
 Already in the original papers on memory burden~\cite{Dvali:2018xpy, Dvali:2018ytn, Dvali:2020wft} it has been argued 
 that black holes must be subjected to this phenomenon. 
 Furthermore, the analysis of~\cite{Dvali:2020wft} revealed
 that the stabilizing nature of the effect, with high likelihood, leads to the prolongation of black hole's life-time. 
  Here, we shall further scrutinize this question in light of
  our analysis.  
      
Let us first structure the arguments that justify
the presence of the memory burden effect in black holes.  
This conclusion can be reached by three independent 
lines of evidence.
  
The first line developed in~\cite{Dvali:2018xpy, Dvali:2020wft}  is based on the universality of the phenomenon.  
The review of the effect given in Chapter~\ref{sec:essenceofmemory}, makes it evident   
that it is impossible to construct a hermitian Hamiltonian
with efficient information storage that avoids 
the memory burden phenomenon.   
The fact that the  black holes 
are efficient (in fact, the most efficient) storing devices, indicates that the effect must be present there. 
 
Further justification is provided by the analysis 
of~\cite{Dvali:2021tez} and by its continuation in the present work. This analysis shows the presence 
of the memory burden effect in solitonic saturons. 
These objects share with black holes all the known key features such as information horizon, area-law entropy, 
thermal decay and Page-like time of information retrieval. 
This gives a strong indication that memory burden
must belong to this list. 
   
Finally, the third line of reasoning is based on 
hints from a microscopic theory, a so-called 
black hole's quantum $N$-portrait~\cite{Dvali:2011aa, Dvali:2012en, Dvali:2013eja}. According to this theory, a black hole represents a saturated coherent state (or a condensate) of gravitons, at criticality.
The term {\it criticality} implies that the mean occupation number is inverse of 
the gravitational coupling. In the present language, these coherent gravitons play the role of a master mode.  
  This theory offers a microscopic explanation of black hole properties within the domain  of calculability. 
  
 In particular, the entropy of a black hole is explained 
 by the emergence of $M$ ``flavors" of the gapless modes  in the above critical state of gravitons.  
  These can be described as Bogoliubov/Goldstone modes of the critical graviton condensate~\cite{Dvali:2012en, Dvali:2015ywa,  Dvali:2015wca,Averin:2016hhm}.
  
   In the present language,
they represent the memory modes.  
   In our reasoning of establishing the nature of 
 the memory burden in black holes, we shall borrow only 
  very general features of $N$-portrait, as this was done 
  in~\cite{Dvali:2018xpy, Dvali:2020wft}. 
   We shall stay strictly within the calculability domain of 
   the QFT framework. 
  
  Namely, we shall rely on the following rather conservative (and natural) starting points:  \\
     
  {\it 1)} Both memory and master modes of a black hole are 
 describable as modes of the graviton field. Although additional field species can certainly exist, e.g., 
   in form of string excitations, they do not change the essence of the effect (although, can lead to
 certain modifications~\cite{Dvali:2021bsy}). We come to the role
of the extra species later, in Subsec.~\ref{subsec:extraspecies}. \\
   
  {\it 2)} The memory modes can be classified according to 
 the symmetries that are left  unbroken  by the black hole metric.   This is analogous to how in our bubble example 
the Goldstone memory modes have been characterized by 
their quantum numbers under the $SU(N-1)\times U(1)$
unbroken symmetry. \\
     
  Equipped with the above guidelines, we shall discuss the black hole memory burden effect. Let us first identify the master and memory modes. 
  
  The identification of the master mode is easy, since 
  at the initial times, this mode has macroscopic 
  occupation number and is therefore well-described classically. 
   Moreover, since the Bogoliubov approximation works for such modes, as a control parameter we can use the 
   suitable characteristics of the classical metric. Basically, the master modes 
   are Fourier harmonics of the classical metric. 
    For a black hole of radius $R$, the dominant 
    contribution comes from the modes of energy gap,  
  \begin{equation} \label{MmsBH}
   m_{\phi} \sim 1/R\,. 
  \end{equation} 
   Of course, the black hole metric viewed as the coherent state of gravitons 
   is a distribution peaked around the above value~\cite{Dvali:2011aa}. For our purposes, it suffices  to consider a single mode rather than a sharply peaked distribution. 
   
         Let us now identify the memory modes.  As already said, 
     we shall only use very general points outlined above. 
  We know that the memory modes must come from 
 graviton modes and can  be classified according to symmetries of the classical metric. Moreover, the number of their flavors, 
 $M$, must be
 \begin{equation} \label{MBH} 
 M \sim  S_{\rm BH}  \,,
 \end{equation} 
  where $S_{\rm BH}$ is the Bekenstein-Hawking entropy which 
  for a black hole of mass $M_{\rm BH}$ and radius 
  $R$ is~\cite{Bekenstein:1973ur}, 
   \begin{equation} \label{SBH} 
 S_{\rm BH} = \pi (R M_{\rm P})^2  = 4\pi \frac{M_{\rm BH}^2}{M_{\rm P}^2} \,.
 \end{equation} 
  
 There exist unique candidates fulfilling the above requirements
 that can emerge as the assisted gapless modes~\cite{Dvali:2017nis, Dvali:2018xpy}. 
 These are the modes of the graviton corresponding to various  spherical harmonics, $Y_{lm}$. 
 Only the harmonics with momenta up to cutoff of the theory must be included, since only such modes are the legitimate 
 (weakly-coupled) QFT degrees of freedom.  
 In Einstein gravity, with no additional light species besides graviton, the cutoff is $M_{\rm P}$. We thus need to include 
 all possible spherical harmonics of graviton up to the Planck 
 mass.  
    
  For a black hole of radius 
 $R$, this counting gives 
 the multiplicity of memory flavors that scales as the area of a black hole in 
 units of $M_{\rm P}$, 
 \begin{equation} \label{MY}
 M \sim  (R M_{\rm P})^2 \,,
 \end{equation} 
 and therefore precisely 
 matches the entropy demand (\ref{SBH})~\cite{Dvali:2017nis, Dvali:2018xpy}.
 
  As an additional supporting evidence,  the above  fully matches the counting of gapless modes from the black hole symmetries derived in~\cite{Averin:2016ybl, Averin:2016hhm}.  

Of course, the majority of modes comes from the highest spherical harmonics. 
  We thus identify the gapless memory modes of a black hole as 
  the modes of angular momenta $\sim M_{\rm P}R$.  
   Their counterparts in the asymptotic vacuum are the 
 same angular harmonics $Y_{lm}$ of a free graviton. 
 However, these carry the energy gaps $m_j  \sim M_{\rm P}$.
 
   Notice that the above explains why a black hole cannot emit information efficiently~\cite{Dvali:2018xpy}: 
in order to ``escape"  from a black hole via a quantum process,
a memory mode has to climb an extremely high 
energetic barrier.     
  
The existence of modes with the same $Y_{lm}$ but with  largely split energy gaps  inside and outside of a black hole is not surprising.  The black hole breaks the Poincare symmetry at the scale $M_{\rm P}$~\cite{Dvali:2020wqi}. Due to this, the memory modes of a black hole, 
 despite having the high orbital momenta, are gapless, whereas the asymptotic modes with similar momenta 
 are gapped by $\sim M_{\rm P}$. 
  
  The above knowledge suffices for adapting the 
  generic Hamiltonian (\ref{Hint}) to a black hole situation. 
  The effective Hamiltonian of black hole memory and master modes is described by (\ref{Hint}) with the 
   understanding that index $j$ labels spherical harmonics $Y_{lm}$. 
   Correspondingly, the intrinsic gaps of master and memory modes are:
   \begin{equation} \label{mphiBH}
   m_{\phi} \sim \frac{1}{R}\, ~{\rm and}~ \,  m_j \sim M_{\rm P}\,, 
  \end{equation} 
 respectively. 
 The coupling of the master mode to memory modes is set by the standard gravitational coupling $\alpha_{\rm gr}(R)$ evaluated at scale $R$, 
  \begin{equation}  \label{NphiBH}
  \frac{1}{N_{\phi}} \sim  \alpha_{\rm gr}(R) \sim  \frac{1}{(RM_{\rm P})^2} \,.
  \end{equation}
  As pointed out in~\cite{Dvali:2011aa}, this coupling is equal to the inverse of the 
 Bekenstein-Hawking entropy $S_{\rm BH}$.  This is not an accident. As explained in~\cite{Dvali:2020wqi},
 this equality puts $S_{\rm BH}$ in accordance with the generic bound on entropy (\ref{AlphaSbound}).
     
   Now, in this description,  the initial state of a classical black hole corresponds to the critical state 
   $n_{\phi} = N_{\phi}$ in which the
  memory modes are essentially gapless, $\omega_j \, = \,0$.  
     
As already discussed in~\cite{Dvali:2018xpy}, this 
explains why at the initial stages of evaporation, the black hole information cannot come out. Indeed, due to the conservation of the angular momentum, translating the information stored in the memory modes into the asymptotic quanta, would require radiating the quanta of energies $\sim M_{\rm P}$, which is not possible. 
In addition, a pair-wise annihilation of memory modes 
of frequencies $\omega$ into the soft (low angular momentum) external quanta is suppressed as 
$\sim \omega_j^5/M_{\rm P}^4$ and is negligible (see below). 
Thus, neither the emission of the memory modes nor their 
conversion into the external quanta is an option.

At the same time, black hole can emit the master mode via quantum scattering, reproducing  the ordinary Hawking radiation~\cite{Hawking:1975vcx}. Basically, the process is a conversion of a master mode into asymptotic quanta of similar frequencies.   
This process is unsuppressed, since the occupation number of the master mode is equal to its inverse coupling. 
As a result, the suppression by powers of coupling is 
compensated by the occupation number,   
and the rate of emission 
is~\cite{Dvali:2011aa},
     \begin{equation} \label{Hawking}
      \Gamma_{\rm ms} \sim \frac{1}{R} \frac{n_{\phi}^2}{N_{\phi}^2}  
     \sim  \frac{1}{R} \,,   
     \end{equation}
   which reproduces the Hawking rate.     
  This simple quantum explanation of the Hawking effect is one of the successes of $N$-portrait~\cite{Dvali:2011aa}. However, due to generic nature of the effect, it holds 
for arbitrary saturons~\cite{Dvali:2021rlf, Dvali:2021tez}.    
    
Since we start in the critical state $n_{\phi} \, = \,N_{\phi}$, at the initial stages 
 the memory burden effect is weak. 
   At this stage, the equation (\ref{Hawking}) is a good approximation. It tells us that, on average, 
  $n_{\phi}$ decreases by $\Delta n_{\phi} \sim 1$ 
 over time $\Delta t \sim R$.  
 
      As it is obvious from (\ref{Hint}), in general, decreasing the occupation number of the master mode by $\Delta n_{\phi}$,  affects its gap by, 
 \begin{equation}     
    \Delta m_{\phi} \sim \frac{q}{N_{\phi}}
    \left (\frac{\Delta n_{\phi}}{N_{\phi}} \right )^{q-1} E_p\,.
   \end{equation} 
   Taking into account (\ref{mphiBH}),  (\ref{NphiBH}) and $N_{\rm G} \sim M \sim S_{\rm BH}$, this is of order, 
   \begin{equation} \label{DeltaM1}
  \Delta m_{\phi} \sim q  \left (\frac{\Delta n_{\phi}}{N_{\phi}} \right )^{q-1} M_{\rm P}\,.
  \end{equation} 
 The memory burden effect becomes dominant 
 by the time when $\Delta m_{\phi} \sim 1/R$, which gives, 
  \begin{equation} \label{DNNBH}    
  \frac{\Delta n_{\phi}}{N_{\phi}} = (q M_{\rm P} R)^{-\frac{1}{q-1}} \,.
   \end{equation}
  The same expression is obtained  from (\ref{minN2}) after  
 taking into account (\ref{mphiBH}).  
 
 The effective gaps of the memory modes also grow 
 with $\Delta n_{\phi}$ as 
    \begin{equation} \label{MemBHgrow}
   \omega_j \, = \, \left (\frac{\Delta n_{\phi}}{N_{\phi}} \right )^{q} M_{\rm P}\,,
  \end{equation} 
 and  by the time of validity of (\ref{DNNBH}), they 
 become, 
 \begin{equation} \label{OmegaBH}    
  \omega_j \, = \, \frac{1}{qR}\frac{1}{(q M_{\rm P} R)^{\frac{1}{q-1}}} \,.
   \end{equation}
  In the above expressions, the quantities  
  $R$ and $S_{\rm BH}$ must be understood as the parameters of the initial black hole. 
  
 We must remember that we track the evolution of an 
 initial state with a given memory pattern. 
 The microstate entropy $S_{\rm BH}$ accounts for the degeneracy
 of various patterns.  However, as the evolution goes on and memory modes gain energy gaps, the initially degenerate memory patterns get split in energy. Of course, the total energies  
 of the systems, i.e., black hole plus radiation, obtained 
 by evolving different initial patterns, 
 are degenerate.
 However, the fractions of energy that remain stored in the black hole relative to radiated portion are different for 
 states evolved from different initial patterns
 (see Fig.~\ref{fig:schematic_memory}).  
 
  The equations  (\ref{DNNBH}) and 
 (\ref{OmegaBH}) determine the onset of the memory burden 
 using as a clock the change in occupation number 
 of the master mode $\Delta n_{\phi}$.   Since before the
 onset of the  memory burden effect, this number changes by usual
 Hawking emission, essentially it measures the time in units 
 of $R$:  $t =  \Delta n_{\phi}  R$.
 
 It also measures the onset of memory burden 
  in terms of the emitted 
 fraction of the initial mass of a black hole as 
 $\Delta M_{\rm BH}/M_{\rm BH} =  \Delta n_{\phi}/N_{\phi}$.
 
Equations  (\ref{DNNBH}) and 
 (\ref{OmegaBH}) depend on an unknown parameter $q$. However, independently of the precise value of this parameter, the tendency is very clear: the domination of the memory burden takes place latest by  the time the 
 black hole radiates away about half  of its mass. 
 That is, the expected upper bound on memory burden time of a black hole is 
 \begin{equation} \label{tMMM}  
  t_{\rm M} \, \sim \, S_{\rm BH} R  \,.  
  \end{equation}       
  This is the case for relatively large $q$. 
    For smaller values of $q$, the memory burden can dominate 
    sooner. For example, for $q=2$, this takes place 
  for $\Delta n_{\phi} \sim \sqrt{S_{\rm BH}}$, i.e., 
  after the time $ t_{\rm M} \sim \sqrt{S_{\rm BH}} R$.   
        
Since the intrinsic gaps of the memory and master 
modes are very different - $m_{\phi}/m_j \sim 1/(M_{\rm P}R)$ - the memory burden effect in black holes is type-$II$.
  
 The coefficient of memory efficiency for a black hole can be estimated following the reasoning of~\cite{Dvali:2018xpy}. 
  In a black hole of mass $M_{\rm BH}$,   
    the energy difference between the most distant information patterns is $\sim 1/R$.   
 If we would store the same pattern by a non-gravitational 
 device of size $R$ on top of a flat space vacuum,  the energy cost would be of order 
  $E_p \sim M_{\rm p} S_{\rm BH}$.  
  The memory-efficiency coefficient at the beginning 
  of evaporation, when black hole is still classical,  is therefore 
  \begin{equation}
   \epsilon_{\rm initial} \sim  \frac{1}{(M_{\rm P}R) S_{\rm BH}} \sim    
   \frac{1}{S_{\rm BH}^{\frac{3}{2}}} \,.
     \end{equation} 
  By the onset of the memory burden effect this becomes, 
   \begin{equation}
   \epsilon_{\rm MB} \sim \frac{1}{RM_{\rm P}} 
   \sim \frac{1}{S_{\rm BH}^{\frac{1}{2}}}
   \,. 
  \end{equation} 
  In the next section we discuss the mass-splitting
  among the stabilized black hole states. 
  
  \subsection{Spread of black hole masses}
   
   Since the mass of a stabilized black hole is a function of the 
   memory burden $E_p$, the prediction is that initially-degenerate black holes over time become spread in masses, as  
    schematically given in Fig.~\ref{fig:schematic_memory}.
  This is similar to the evolution of equal mass vacuum
  bubbles with different values of $N_{\rm G}$. 
       
  In order to share with the reader a general sense of scaling, 
let us analyse an oversimplified toy model in 
which the memory modes of a black hole are treated as 
independent qubits with the Hamiltonian (\ref{Hint}) with 
equal intrinsic gaps $m_j = M_{\rm P}$. 
In this case, $E_p = M_{\rm P} N_{\rm G}$. Correspondingly, the memory burden effect can be 
measured by the total occupation number of the memory modes $N_{\rm G}$.
For $M$ qubits, this number is bounded by 
  $N_{\rm G} \leqslant M$ and the total number of patterns 
  is given by  $n_{\rm st} = 2^M$.  This gives the entropy 
  $S_{\rm BH} \sim M$ (more precisely, $S_{\rm BH} = M\ln2$).  

 For estimating the spread in this toy model 
 we can use the equation (\ref{DeltaM1}) and take into account that
 the memory burden effect becomes significant for $\Delta m_{\phi} \sim 1/R$. We also take into account that for a black hole 
$m_{\phi} \sim 1/R,~  m_j \sim M_{\rm P} \sim M$ and $N_{\phi} \sim M \sim S_{\rm BH}$. 
 However, we keep $N_{\rm G}$ as a free parameter.  
 
  Correspondingly,  for a black hole of the initial mass
 $M_{\rm BH}$ and the memory burden $N_{\rm G}$, 
 the relative change in mass by the time of stabilization 
 is, 
  \begin{equation} \label{DMBHMBH}
\frac{\Delta M_{\rm BH}}{M_{\rm BH}} \, \sim \, 
 \left( \frac{S_{\rm BH}}{q(M_{\rm P}R) N_{\rm G}} \right )^\frac{1}{q - 1} \,\sim \, \left( \frac{\sqrt{S_{\rm BH}}}{q N_{\rm G}} \right )^\frac{1}{q - 1} \,. 
\end{equation} 
 One must remember that the above expression is valid 
 as long as 
 \begin{equation} \label{NGbound}
  N_{\rm G} \geqslant \frac{1}{q}\sqrt{S_{\rm BH}} \,.
 \end{equation}
  This inequality represents a manifestation of the general formula
(\ref{ElargerEstar}) for a black hole.  In the opposite case the stabilization is not efficient. 
 Thus, for $N_{\rm G}$ at its lower bound (\ref{NGbound}), the  
 stabilization takes place at $\Delta M_{\rm BH}/M_{\rm BH} \sim 1$. 
On the other hand, for $N_{\rm G} \sim S_{\rm BH}$, we 
 recover (\ref{DNNBH}).
  
The equation (\ref{DMBHMBH}) tells us that the spread in masses of stabilized remnants is determined by the statistical distribution of $N_{\rm G}$ among the initial black holes.   
  For simplicity of estimate,  let us assume that, at the time of formation,
 the occupation number of the black hole master mode is critical, $n_{\phi} = N_{\phi}$. That is, the memory patterns are strictly degenerate. In practice, this is expected to be 
an extremely good approximation, since the initial black hole 
is well-described classically up to corrections 
$\sim 1/N_{\phi}$.  
 
 In such a case, we can assume that the probability distribution
of patterns is flat, with no energy bias.  Correspondingly, the probability of a pattern with given  $N_{\rm G}$ is, 
 \begin{equation} \label{PofNG}
    {\mathcal P}_{N_{\rm G}} \,  = \, 2^{-M} \frac{M!}{(M-N_{\rm G})! N_{\rm G}!} \,, 
 \end{equation}  
which is maximal for $N_{\rm G}  = M/2$ with 
  \begin{equation}
    {\mathcal P}_{M/2} \,  \sim  \frac{1}{\sqrt{M}} \,,
 \end{equation}    
 and the width of $\sqrt{M}/2$.
On the other hand, for $N_{\rm G} \ll M$, the probability is exponentially suppressed as,   
  \begin{equation}
    {\mathcal P}_{N_{\rm G} \ll M} \,  \sim  2^{-M} \left( \frac{M {\rm e}}{N_{\rm G}} \right 
  )^{N_{\rm G}}    \,,
 \end{equation}
 where we dropped polynomial factors for simplicity. 
  Taking into account that the black hole entropy 
  is $S_{\rm BH} \sim M$, the above implies that  
the  probability of forming a black hole
 with memory burden $N_{\rm G} \ll S_{\rm BH}$ is exponentially 
 small. Thus, it is expected that the 
 most probable black holes are the ones with 
 $N_{\rm G} \sim S_{\rm BH}$.\\

  To summarize, as it is clear from the equation 
  (\ref{DMBHMBH}), the change of the black hole mass 
 prior to its stabilization is set by $N_{\rm G}$ as, 
 \begin{equation} \label{SpreadDM}
\Delta M_{\rm BH} \, \propto \,  N_{\rm G}^{-\frac{1}{q-1}}\,,
 \end{equation}
 where the coefficient of proportionality,  
$M_{\rm BH}\, 
  \left( \sqrt{S_{\rm BH}}/q \right )^\frac{1}{q - 1}$,  
  is fully determined by the initial mass of a black hole.  
  At the same time, the memory burden $N_{\rm G}$ is determined by the mass only statistically, via (\ref{PofNG}).
    
 This fact has potentially important observational 
 implications since it predicts a statistical spread 
 of $\Delta M_{\rm BH}$ in addition to 
 the initial mass distribution determined 
 by particularities of a cosmological scenario (see Sec.~\ref{sec:implications}).  
 
\subsection{Fate of a black hole burdened by memory}
     
 What happens after a black hole enters the memory burden phase 
requires a more detailed understanding of the picture. The two possible outcomes were discussed in~\cite{Dvali:2020wft}.  
 
 The first option is that a new classical (collective) 
 instability sets in and the (former) black hole evolves 
 through it.  From our current understanding, it cannot be excluded 
 that due to this instability the remnant 
 can disintegrate via some non-linear process.  

 The second (more conservative) option assumes no immediate classical instability. 
 In such a case, the black hole continues to decay via a quantum
 process. However, due to the memory burden, the process is extremely slow. In~\cite{Dvali:2020wft}, the remaining life-time of a black hole was given as, 
  \begin{equation} \label{tauBHk}
  \tau \sim  R\, S_{\rm BH}^{1+k}\,, 
  \end{equation}
where $k>0$ is an integer. This form follows from the fact 
that the prolonged life-time is an analytic function of $S_{\rm BH}$.
The analyticity in $S_{\rm BH}$ is enforced by the requirement that the
decay rate must be analytic in occupation numbers as well as in gravitational couplings, all of which are set by $S_{\rm BH}$.  The case $k=0$ (zero memory burden) 
would correspond to a standard extrapolation of 
Hawking's decay rate.  

The value $k > 0$, can be understood  
  from the following argument.
   In order to continue its decay, the black hole must get rid 
   of the memory burden. That is, the excited memory modes 
   must get de-excited. This can only be done by the  
   scattering processes that involve at least a pair of 
   the memory modes. The memory modes must annihilate into the 
   modes of lower angular momenta in order to match their energies.  
   That is, each mode $Y_{lm}$ must find a partner 
   $Y_{l'm'}$ with very close values of $l,m$. 
   Such pairs are extremely rare and their 
   annihilation rate is 
   \begin{equation} \label{RateMS}
   \Gamma \sim \frac{\omega_j^5}{M_{\rm P}^4}
   \sim \frac{1}{R^5M_{\rm P}^4}\,.
   \end{equation}  
   In terms of the initial entropy, the life-time translates 
  to 
  \begin{equation} \label{tauBHbound}
  \tau \gtrsim   R\, S_{\rm BH}^{2}\,. 
  \end{equation}
  This reinforces equation (\ref{tauBH}), 
  giving $k=1$ as the most conservative estimate. 
  
 \subsection{The Effect of Species}  \label{subsec:extraspecies}

  It is known that  black holes provide a  
 link between the number of QFT species $N_{\rm sp}$ 
 and the upper bound on the scale of strong quantum gravity~\cite{Dvali:2007hz, Dvali:2007wp, Dvali:2008fd, Dvali:2008ec, Dvali:2009ks, Dvali:2010vm}, 
 \begin{equation} \label{Msp}
   M_{\rm sp} \, = \, \frac{M_{\rm P}}{\sqrt{N_{\rm sp}}} \,. 
 \end{equation}   
  The expression is fully non-perturbative and cannot be 
 removed by any re-summation of perturbative series.  
 The  physical meaning of the scale $M_{\rm sp}$, called ``species scale", is that a black hole of size $< \, 1/M_{\rm sp}$, cannot be treated semi-classically. This can be seen from a number of arguments. \\ 
 
 For example, if the standard 
 semi-classical evaporation rate would apply to such a black hole, one would conclude 
 that the black hole would live shorter than its radius, which is absurd.   
 
 Similarly, the violation of the bound (\ref{Msp})
 is excluded by quantum information arguments~\cite{Dvali:2008ec}: 
 a black hole of size $< 1/M_{\rm sp}$, would have the 
 microstate degeneracy, and correspondingly the information storage capacity, exceeding the one of
Bekenstein-Hawking entropy (\ref{SBH}). 
In other words, the ``species entropy"  would exceed 
the entropy of Bekenstein-Hawking~\cite{Dvali:2020wqi}. 

  In general, the increased number of species 
  shortens the standard stage of black hole evaporation
 by a factor $1/N_{\rm sp}$. This has implication 
 for the memory burden, since the time 
 for reaching this phase is correspondingly shortened~\cite{Dvali:2021bsy}: 
  \begin{equation} \label{tMsp}  
  t_{\rm M} \, \sim \, S_{\rm BH} R \frac{1}{N_{\rm sp}}  \,.  
  \end{equation}  
  The effect of the species beyond this point is less certain,  since due to quantum hair, no universality in decay rates of different 
  species is guaranteed. In other words, the black hole can develop 
  a ``species hair"~\cite{Dvali:2008fd}.
  Therefore, the further decay of the remnant 
 can be biased towards some of the species.
  
   As discussed in~\cite{Alexandre:2024nuo}, 
    already the formula (\ref{tMsp})  
  can have important implications for PBH dark matter, as 
  it can shift the masses of memory burdened 
  black holes towards higher values. Furthermore, 
  assuming that decay of the remnant remains 
  approximately democratic in species, the 
  life-time in memory burden phase will subsequently 
  be shortened by the quantity $\sim 1/N_{\rm sp}$.  
     
\section{Quantum memory burden 
versus  classical extremality} 
\label{sec:extremality}

It is well known that black holes, as well as solitons,
can be stabilized against the quantum decay by {\it classical} charges, either Noether or topological.   
The term  ``classical" implies that the charge 
is classically-detectable. In other words, the object must carry 
a classical ``hair" with respect to this charge. 

In case of a black hole, in the light of
``no-hair" theorems~\cite{Ruffini:1971bza, Bekenstein:1972ny, 
Teitelboim:1972pk}, such are the charges associated 
with massless gauge fields that 
can be measured via Gaussian fluxes at infinity \footnote{
The seeming exceptions, such as a classical skyrmion hair~\cite{Luckock:1986tr, Bizon:1992gb, Droz:1991cx},  
have been shown~\cite{Dvali:2016sac, Dvali:2016mur} to 
be equivalent to Aharonov-Bohm type hair 
under a discrete gauge symmetry~\cite{Krauss:1988zc, Preskill:1990bm}.}. 
Usually, the amount of charge 
capable of stabilization is comparable to the mass of a 
black hole (or a soliton). 

 We now wish to confront a mechanism of stabilization of an object by
 a classical hair with the one of a  
 quantum memory burden.  
 For achieving this, we first need to understand the stabilization of a black hole (or a soliton) by a classical charge in a fully  quantum language of a microscopic theory.   
 
  The first steps in this direction   
 were undertaken in~\cite{Dvali:2011aa}.  
 This microscopic description was further generalized 
 to topological solitons in~\cite{Dvali:2015jxa}. 
 
 In black hole's quantum $N$-portrait~\cite{Dvali:2011aa}, a purely gravitational black hole represents a critical coherent state or a condensate of master mode gravitons. 
 The typical energy gaps of these modes are $m_{\phi} \sim 1/R$ and  
  their mean-occupation number, 
  $n_{\phi}$, is critical, $n_{\phi} \sim  S_{\rm BH}$. 
  
   Correspondingly, as in any other system with enhanced information capacity, prior to stabilization, the master modes are the main contributors into the mass of an initially-classical black hole, 
    \begin{equation}
    M_{\rm BH} \sim n_{\phi}\, m \sim S_{\rm BH} \frac{1}{R} \,.
   \end{equation}
  In this quantum picture, the existence 
  of a classical charge of a black hole means that, 
on top of the master mode, some other mode carrying this  particular charge becomes macroscopically occupied. 
In other words, a second master mode appears. 

  For example, the electrically 
charged  Reissner-Nordstr\"om black hole contains a 
macroscopic occupation number of photons. 
These photons compose a coherent state
describing the classical electric field of the black hole. 
  That is, when a black hole carries a classical charge, 
  the ``status" of a master mode is shared between the 
  master graviton and some other mode.  
 
 Now, as argued in~\cite{Dvali:2011aa},
 the occupation number of the master graviton sets the upper  bound on other occupation numbers.  
That is, for a black hole the occupation numbers of master modes satisfy the bound, 
\begin{equation} \label{AnyMode}
   n_{\rm any~mode}  \leqslant n_{\phi} \,.  
 \end{equation}
  In this way, the theory~\cite{Dvali:2011aa} provides a quantum 
  explanation of why a black hole charge can never exceed its mass.   
Notice that this is true for an arbitrary saturon. In particular,~\eqref{AnyMode} is satisfied by saturated vacuum bubbles
discussed in section \ref{sec:memoryburdensoliton}.

According to~\cite{Dvali:2011aa}, a black hole reaches 
extremality when the occupation number of a certain mode, other than the master graviton,  saturates the above
 bound. In this description, the stability of an extremal 
 Reissner-Nordstr\"om  black hole 
 can be viewed as the saturation of the bound~\eqref{AnyMode}  
 by the photon constituents of a black hole.
 
  It is intuitively clear~\cite{Dvali:2011aa} as well as evident from computations~\cite{Dvali:2022vzz} that in this case the black hole 
  evaporation must stop. The reason is that for maintaining the 
  saturation of the bound \eqref{AnyMode}, a further
 lowering of the occupation number of the 
 graviton master mode, $n_{\phi}$, must be accompanied by 
 the depletion of the electric field. This is not possible 
 without emission of the charged quanta. However, the fact that 
 the black hole of size $R$ has an unscreened electric 
 charge to start with, implies that in the theory the mass 
 of the lightest charged particle, $m_{\rm e}$, 
 satisfies: $m_{\rm e} \gg 1/R$. In the opposite case, the 
 charge would be screened by the Schwinger effect. 
 
 Correspondingly, the quantum emission of electric charge 
 is exponentially suppressed, since such a process 
 requires a rescattering of many soft constituent quanta (of energies $\sim 1/R$) into a quantum of energy $m_{\rm e} \gg 1/R$. 
 The transition rate of such processes is  bounded from above by $e^{-m_{\rm e}R}$, as indicated by a general argument given in~\cite{Dvali:2020wqi}. This is confirmed by 
 explicit computations of corresponding multiparticle
 processes~\cite{Dvali:2022vzz}. 
 The suppression of the transitions many$\rightarrow$few can 
 also be extracted from explicit computations of graviton scattering processes~\cite{Dvali:2014ila, Addazi:2016ksu}. The analogous suppression 
 is exhibited by scalar theories~\cite{Brown:1992ay,  Argyres:1992np, Voloshin:1992rr, Gorsky:1993ix, Libanov:1994ug, Libanov:1995gh, Son:1995wz, Monin:2018cbi}. 
 
   Thus, the characteristics of black hole stabilization by a classical charge is the existence of a long-range classical hair with respect to that charge. At the same time, the quantum information pattern carried by the black hole 
plays no role. 
  For example, the extremal black holes of the same electric charge and mass can carry information patterns of very different content. 
   
   In contrast, when a black hole is stabilized by a
    quantum memory burden, no classical charge associated 
   with a long-range gauge field is required. 
   Of course, a macroscopic ``hair" emerges in form of 
   a memory burden parameter, $N_{\rm G}$. 
   This parameter is in principle measurable by a scattering experiment, but there is no inconsistency with the no-hair 
   properties: since the memory-burdened black hole is a quantum object, it has no reason to respect classical no-hair theorems.  
   Therefore, unlike classical extremal case,  
 there is no specific gauge charge or a long-range Gaussian flux associated with the quantity $N_{\rm G}$.  
 
  The difference between the two 
  mechanisms is also clear from the fact 
  that the decay of a memory-burdened black hole 
  is not suppressed exponentially but only slows down as 
  a power-law \eqref{tauBH}.  
  
  Interestingly, the solitons stabilized by the memory burden 
  unify both features.  
  The main reason is the existence of two types of microstates 
  discussed above.  
  
  The first category of microstates have equal $N_{\rm G}$ but are distinguished  by the
  relative  $SU(N)$-transformations.  These remain exactly degenerate even after the memory burden sets in.  
  The microstates belonging 
  to the second category are distinguished by values of $N_{\rm G}$. 
  The energies of such microstates get split after stabilization. 

Both sets of microstates have counterparts in a 
black hole. 
  As discussed, the microstates of a black hole 
   correspond to degeneracy of patterns obtained by 
  occupation numbers of different memory modes $Y_{lm}$.
   Among these states, there exist both categories: 
   with equal and distinct memory burdens.
   As time elapses, the level splitting develops 
  according to the memory burdens they carry, as given in
  Fig.~\ref{fig:schematic_memory}. 
   
  Another example of a classical ``charge"  that can stabilize a black hole against  quantum decay is the angular momentum. 
  The similar case in solitons  reveals a certain interesting 
 peculiarity.  
        
   The stationary $SU(N)$-bubble with non-zero spin is obtained via giving the winding number $n$ to the Goldstone mode~\cite{Dvali:2021ofp}. Basically,  the matrix $U$ in (\ref{UUU}) is replaced by
  \begin{equation} \label{Uwinding}  
     U = e^{i \left(t\omega + n \varphi/\sqrt{2}\right)  \frac{\omega_j}{\omega} T^j}\,, 
  \end{equation}    
where $\varphi$ is the polar angle and the  factor $\sqrt{2}$ is added to ensure the phase continuity.  
 
 The winding of the Goldstone phase produces the vorticity and, simultaneously, the angular momentum, 
 \begin{equation} \label{Jn}
    J =\int{\rm d}^{3}x\, T_{0\varphi} = n\,N_G = n\,Q \,,
  \end{equation} 
  where $T_{\mu\nu}$ is the energy momentum tensor. 
  
  The fact that the winding number 
  $n$ induces the spin given by~\eqref{Jn}
 has been known for $Q$-balls with $U(1)$-charge~\cite{Kim:1992mm, Volkov:2002aj,Kleihaus:2005me}.  
  The remarkable thing about  the saturated $SU(N)$ vacuum bubbles is 
  that the equation (\ref{Jn}),  describing the relation between the
   maximal spin and the entropy, is strikingly similar to 
  the one satisfied by a black hole~\cite{Dvali:2021ofp}, 
  \begin{equation} \label{JmaxS}
    J_{\rm max} = S \,.
  \end{equation}  
   A saturon vacuum bubble of $SU(N)$, which has entropy  
  \begin{equation} \label{S=NG}
   S =  N_{\rm G}\,,
  \end{equation}   
reproduces the relation \eqref{JmaxS} due to the limited vorticity, $n\sim 1$, that it can sustain. Attaining a higher spin would require a larger vorticity, $n \gg 1$, which would increase the mass and the size of the bubble, thereby making it undersaturated. 

 Note that the relation (\ref{JmaxS}) is shared by other saturons. For example, it is automatic 
in saturated baryons in  QCD with large number of colors and flavors~\cite{Dvali:2019jjw}. 

The same relation is shown~\cite{Dvali:2024xyz} to be shared by a saturated version of a spinning cosmic string loop ( a so-called vorton~\cite{Davis:1988ij}). 

 To summarize, the saturon bubbles of maximal spin 
 exhibit the exact same relation between spin and entropy 
 as the extremal spinning black holes. At the same time, 
 in extremal saturon bubbles this relation reveals an explicit 
 microscopic meaning in terms of vorticity.  
    
   Notice also that vorticity gives a topological meaning to the 
   quantum stability of the extremal bubble~\cite{Dvali:2021ofp}. 
   Indeed, the decay of a bubble via the particle emission 
   is blocked because of conservation of the topological winding number. The decay is only possible if the entire vortex is ejected via a quantum tunnelling which is an exponentially suppressed process. 
      
  In this sense, the extremal spin represents another form 
 of a ``classical" memory burden. 
 As in the case of the burden induced by a 
 pure $SU(N)$-charge, the burden induced by the
 spin is due to a critical occupation number $N_{\rm G}$ 
  of a Goldstone/memory  mode. For a saturon bubble, 
  $\omega \sim m$, this memory mode becomes an additional master mode.   
   The difference from non-spinning case is that 
 the macroscopically-occupied memory mode has a non-zero angular momentum. 
 
  This analogy is fully extended to black holes if we 
 conjecture that a spinning black hole corresponds to the 
 one in which a mode of a particular angular harmonic $Y_{lm}$ is macroscopically occupied. In this light, it is interesting to discuss  
  a conjecture of~\cite{Dvali:2021ofp} stating that the extremal spin in a black hole is accompanied by vorticity,  
 similarly to the case of a spinning saturon bubble.  
  Again, as in the case of a bubble, the vorticity gives 
   a topological explanation to the quantum stability
   of the extremal black hole.  
   In the case of the spin, similarly to
   other classical charges, the extremal black hole does carry a classical hair in form of the angular momentum. 
  
 To summarize, there exist fundamental differences
 between the  black holes stabilized  via 
 classical charges and the ones stabilized by a quantum memory burden effect. 
  One obvious difference  is  that the classical 
extremality is necessarily accompanied by a long-range classical gauge hair. 
 
  This is not the case 
 for a black hole stabilized by the memory burden effect. 
 Instead, for a memory-burdened black hole, the hair is short-range and carries no Gaussian flux at infinity. 
 
More importantly, a memory-burdened black hole carries a significant quantum hair. The significance is measured by the fact that the energy of the quantum information pattern is 
comparable to the mass of the remnant.  

 Such a remnant is neither a black hole nor classical,
 and, therefore, is not constrained by classical no-hair theorems. 
 Although macroscopic, it nevertheless is a quantum entity.  
 This ``macro-quantumness" is a feature of 
 black holes~\cite{Dvali:2013eja} as well as of 
 other saturons~\cite{Dvali:2020wqi}.  \\

The above important difference between the quantum memory-burdened 
and (quasi) extremal black holes must be taken into account when applying such mechanisms of stabilization to light PBH dark matter. In particular, if the naive semi-classical time-scale required for approaching extremality is longer than the memory burden 
time $t_{\rm M}$, the latter effect becomes the dominant one and 
must be taken into account. 

 In addition to differences at a fundamental level, this will have observational consequences. 
 For example, the memory-burdened PBH dark matter~\cite{Dvali:2020wft, Dvali:2021byy, Alexandre:2024nuo, Thoss:2024hsr, Balaji:2024hpu, Haque:2024eyh}  will be subjected to the mass-spread
 discussed in the present paper.  This spread is absent 
 for PBHs stabilized by extremality within the validity of 
 semi-classical regimes, such as discussed in~\cite{deFreitasPacheco:2023hpb} (and references therein). \\

   The fundamental difference between the prolongations 
of a black hole life-time due to a classical hair versus the quantum memory burden effect, extends beyond the ordinary black holes.  The example is provided by a black hole smaller than the compactification radius in the framework of large extra dimensions~\cite{ADD}. 
 
  Such a black hole is a solution of Einstein's equations 
 in high-dimensional gravity and its  radius and temperature are defined by the fundamental Planck mass, $M_{\rm F}$, which is 
 suppressed relative to the four-dimensional one by the volume 
 of extra space~\cite{ADD}: 
 \begin{equation}  \label{ADD1} 
   M_{\rm F} \, = \, \frac{M_{\rm P}}{\sqrt{V_{\rm extra}}} \,.
 \end{equation} 
  Correspondingly, the  high-dimensional black holes are colder relative 
 to would-be four-dimensional black holes of the same mass. 
  It was proposed in~\cite{Anchordoqui:2024dxu,Anchordoqui:2022txe} that such black holes can serve as dark matter. 
    
   Let us confront the Kaluza-Klein (KK) hair of {\it young}  high-dimensional black holes 
   with the memory burden effect. Of course, the question arises 
   solely from the point of view of a four-dimensional observer, since 
   from high-dimensional perspective these are usual classical black holes. 
      
      From four-dimensional perspective, the decay of high-dimensional black holes can be understood in terms of species.
    First, as discussed in~\cite{Dvali:2007hz}, the equation (\ref{ADD1}) can be viewed as a particular case  of (\ref{Msp}), with the role of particle species played by KK gravitons  and the role of the
    four-dimensional species scale assumed by the fundamental Planck mass $M_{\rm sp} = M_{\rm F}$. 
    
    When evaluating the evaporation rate, we  must take into account that the black hole is smaller than the compactification radius, correspondingly, it develops the 
    {\it species hair}~\cite{Dvali:2008fd}.
    That is, the black hole sources a tower of massive KK gravitons to which  it also evaporates. 
 The evaporation into many KK species is equivalent to the emission of a single  high-dimensional graviton~\cite{Dvali:2009ks}.  

   Therefore, the evaporation of a young  black hole is well-described semi-classically through a high-dimensional Hawking 
process. From the point of view of a four-dimensional 
observer it is an object that caries a classical species hair~\cite{Dvali:2008fd} under KK gravitons. Such a classical hair is fundamentally different from  the quantum  memory burden that sets in at a later stage. 
 
   This is already evident from the fact 
 that quantum memory burden will be experienced 
 by a high dimensional black hole after it gets sufficiently  
 ``old".  Extrapolating our results, we can say that 
 this will happen the latest by half-decay. 
 Therefore, for a high-dimensional PBH, the quantum memory burden 
 effect must be taken into account in the same way as for 
 ordinary Einsteinian black holes. 

\section{Numerical Analysis: stabilization of solitons}\label{sec:numerics}

We summarize here our numerical findings. We focus on the dynamics of underburdened solitons, whose evolution, as discussed in Sec.~\ref{sec:extremality}, offers analogy with evaporating black holes that are stabilised by their quantum memory.

   The idea is to start with a vacuum bubble,  
   with the value of the charge $N_{\rm G}$ given by \eqref{ECV}.
   As discussed in Subsec.~\ref{subsection:stabilizationbymemory},
   this amount of charge  is insufficient for  an immediate stabilization. 
   In particular, for such a bubble the  energy of the memory pattern 
   is negligible as compared to the energy of the master mode 
   \eqref{ECV}, which constitutes the main source of the energy.  
   
    Such a bubble oscillates, emitting energy and correspondingly 
shrinking in size. We map this stage of evolution on the 
one of a black hole before entering the memory burden phase. 
    
   The decay process of the bubble continues up until the Goldstone charge fulfils the condition \eqref{QNG}. Simultaneously, the 
   energy of the memory pattern catches up with the energy of the master 
   mode \eqref{EC}. 
   Correspondingly, the bubble 
   is stabilized by the burden of memory. At this point, the bubble becomes a stationary $Q$-ball.  
    
Let us denote the reference stationary $Q$-ball charge by $Q_{\rm s} = N_{\rm G}$ and its frequency by $\omega_{\rm s}$. We then vary the initial charge by changing the frequency of constituents, while keeping the profile fixed, to the following values
\begin{itemize}
    \item $\omega = 3\omega_{\rm s}/4$, $Q=3Q_{\rm s}/4$
    \item $\omega = \omega_{\rm s}/2$, $Q=Q_{\rm s}/2$
    \item $\omega = \omega_{\rm s}/4$, $Q=Q_{\rm s}/4$
    \item $\omega = 0$, $Q=0$.
\end{itemize} 
resulting in underburdened configurations of different $N_{\rm G}$ (therefore, not related by a $SU(N)$ transformation) in the regime~\eqref{ECV} and \eqref{NGV}.

Our numerical analysis supports the analytical discussion of the previous Sections. Namely, the bubbles of different initial charges - but of similar energies - while they have analogous dynamics at initial times, are stabilised towards $Q$-ball configurations of different masses.

The statement holds true for all regimes analysed in our study - regardless of other details such as the winding number, the inclusion of new interactions and the thickness of the bubble wall. This shows that the mechanism of stabilization by memory is a general consequence of the high-memory storage capacity of the object and is only secondarily affected by other features, such as, for example, saturation or vorticity. Consequently, this further supports the prediction of Subsec.~\ref{subsec:energysplitting} that an analogous mass splitting is expected in the asymptotic distribution of black holes of equal initial masses but different memory burdens.

The rest of this Section is dedicated to showing how the above statement comes about. For numerical simplicity, we focus on the $SU(4)$ symmetric case. Unless otherwise stated, we fix $\alpha = m=1$ in units of $f$ and choose $\omega=0.3\, m$. Since we are interested in 
the winding $n=0$ and $n=1$ cases, a $2+1$ dimensional analysis is sufficient. Finally, some of the simulation visuals can be found at the following \href{https://youtu.be/boDpRXJnT5E}{URL}.

\subsection*{Winding $n=0$ case}
\begin{figure*}[th!]
    \centering
    \includegraphics[width=1\textwidth]{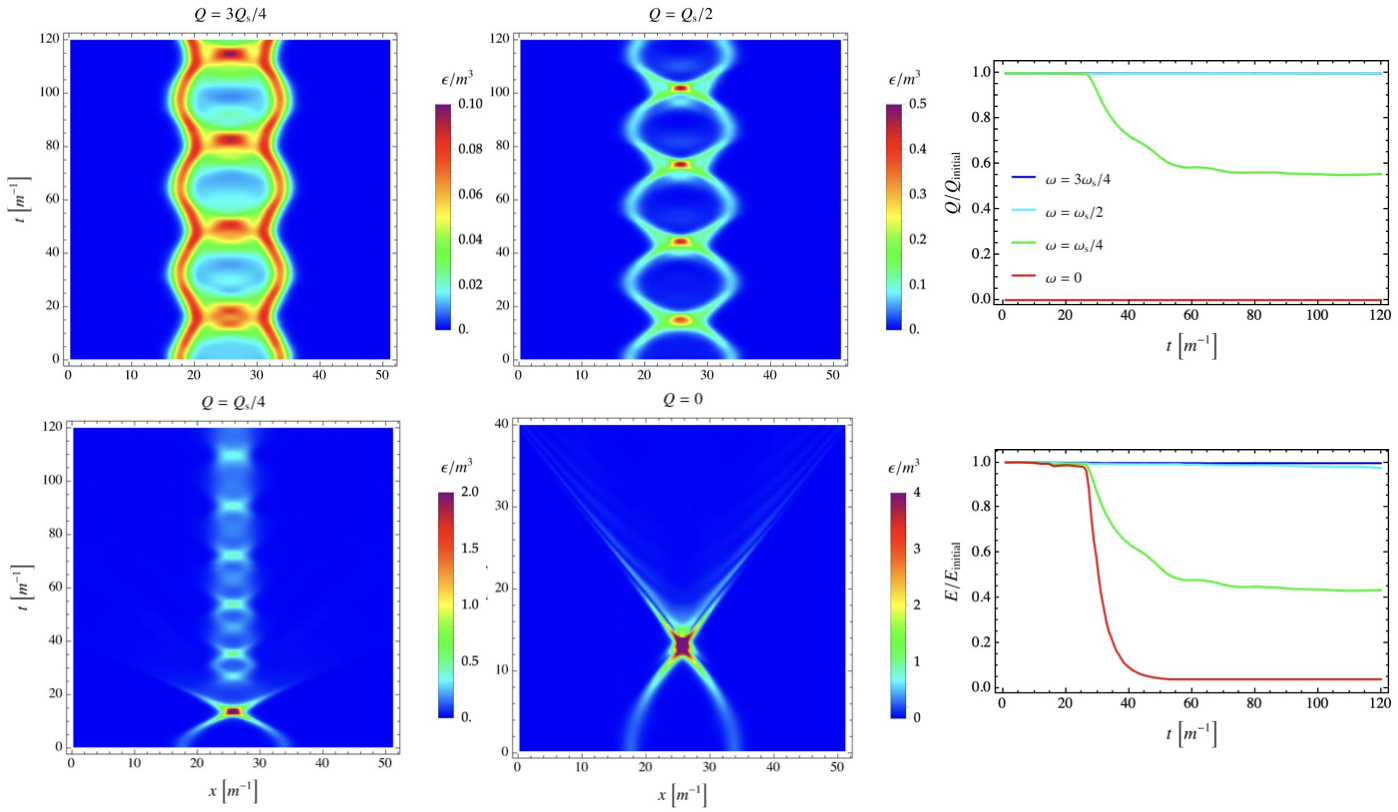}
    \caption{Energy density evolution for different initial charges and winding $n=0$. The third column shows the integrated charge and energy as functions of time.}
    \label{fig:energy_n=0}
\end{figure*}

The energy density evolution is shown in Fig.~\ref{fig:energy_n=0}. For $Q=3Q_{\rm s}/4$ (first panel), the system is close to the stationary case, and therefore has a significant burden. In fact, the bubble starts collapsing, but the information (charge) stored within immediately backreacts on the dynamics. This, in turn, results in an oscillatory behaviour analogous to what was already observed in~\cite{Dvali:2021tez}. The reason the bubble does not relax to its stationary configuration is energetic. Simply, the excitations are smaller than the mass gap outside of the configuration. As a consequence, (almost) no charge is emitted and the information is retained within the bubble. 
This is a manifestation of the existence of the information horizon~\cite{Dvali:2021tez}.  The notion becomes exact in 
the semi-classical limit which corresponds to large-$N$.

This can be seen explicitly in the third column of Fig.~\ref{fig:energy_n=0} where the integrated energy and charge densities are shown as functions of time (the integrated charge, corresponding to the blue line, is obscured by the cyan curve of the case $Q=Q_{\rm s}/2$). 

An analogous situation is observed in the second panel of Fig.~\ref{fig:energy_n=0}. Since in this case the initial charge is smaller - $Q=Q_{\rm s}/2$ - the resulting oscillations are more pronounced. Still, the burden forbids the collapse, and the energy gap ensures that almost no charge is emitted. The oscillatory frequency of the profile in both cases is roughly given by $R^{-1}$ as it can be qualitatively seen from the plot. A perturbative analysis of the system energy would inform us of this fact, as already shown in~\cite{Dvali:2021tez}.

In the first panel of the second row, for $Q=Q_{\rm s}/4$, the energy of the collapse is sufficient for overcoming the mass gap outside of the $Q$-ball. As a consequence, charge is emitted and part of the information is lost in the first oscillation as it can be seen in Fig.~\ref{fig:energy_n=0}. At later stages, the pulsation energy becomes  insufficient for exciting new quanta, resulting in a stationary configuration with smaller charge and radius. For analysis of such excitations in $U(1)$ symmetric $Q$-balls, see e.g.,~\cite{Saffin:2022tub}.

Finally, in the zero charge no burden is present to stop the collapse. The system is effectively a bubble interpolating between two degenerate vacua.  Therefore, the collapse due to the surface tension takes place within one oscillation and is rather violent. However, for smaller values of coupling, the oscillation can last longer. 

\subsection*{Winding $n=1$ case}\label{subsec:winding1}
 \begin{figure*}[th!]
    \centering
    \includegraphics[width=.98\textwidth]{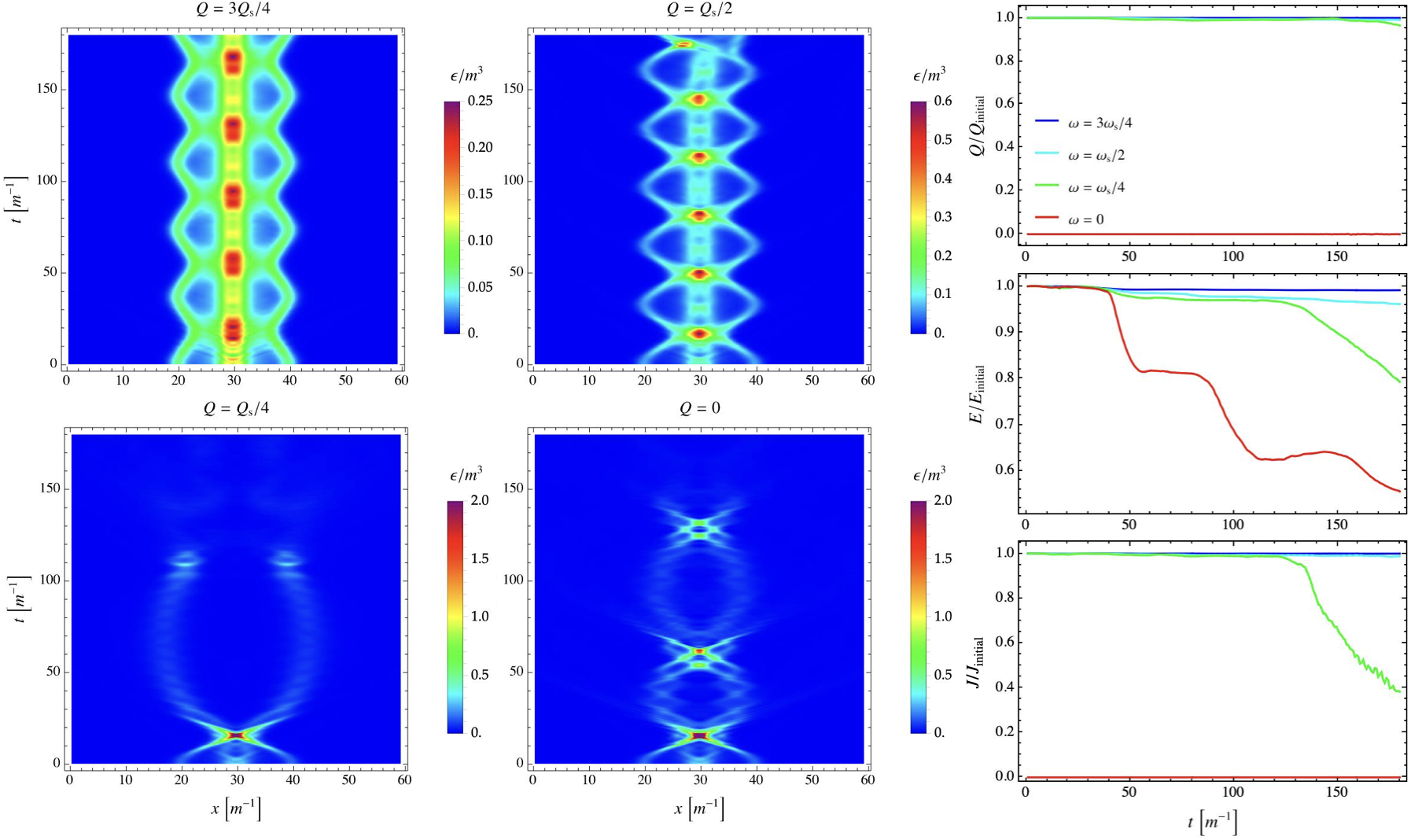}
    \caption{Energy density evolution for different initial charges for winding number $n=1$. The third column shows the integrated charge and energy and angular momentum as functions of time}
    \label{fig:energy_n=1}
\end{figure*}
Memory burden has an analogous stabilizing impact on underburdened bubbles endowed with vorticity characterized by winding number $n=1$. We refer the reader to~\cite{Zantedeschi:2022czs} for the numerical construction of the stationary solution.

As discussed in Sec.~\ref{sec:extremality}, vorticity is realized in terms of a macroscopic occupation number of Goldstones under the ansatz \eqref{Uwinding}, leading to non-vanishing angular momentum according to \eqref{Jn}.  

Since saturated bubbles with vorticity and black holes obey the same maximal-spin bound - c.f., \eqref{JmaxS} - one might be tempted to map underburdened bubbles with vorticity on evaporating black holes dynamically approaching extremality. Of course, within the semi-classical picture, whether or not a black hole evolves towards such point, is determined by the particle spectrum participating in the emission~\cite{Page:1976ki}. However, the semi-classical analysis 
does not take into account the memory burden effect which 
invalidates it. This must be kept in mind when 
mapping the quantum evolution of a black hole 
on the dynamics of underburdened bubble in our numerical analysis. 

The first panel in Fig.~\ref{fig:energy_n=1} shows the evolution of energy as a function of time for the case $Q=3Q_{\rm s}/4$. Although the simulation is performed in two spatial dimension, for illustrative purposes, we slice here along the $x$ coordinate, fixing $y$ along the center of the vortex. The situation is similar to the $n=0$ case with similar initial charge. Notice that in this case the energy of the vortex in the central region oscillates together with the pulsating radius of the configuration. The period of pulsation also here is of order $R^{-1}$ - $R$ being the radius of the stationary configuration (and therefore, also the initial radius at the simulation time). As expected, the bubble is stabilized by the conserved charge and the angular momentum in its interior - see third column of Fig.~\ref{fig:energy_n=1}.

A similar behaviour is observed in the second panel for $Q=Q_{\rm s}/2$. Indeed here the pulsations are more pronounced due to the larger initial off-balance between memory and master mode energy. Concomitantly, a mild emission of energy and charge is observed through the initial collapse. 
Notice that at the end of the simulation, at large time, the axial symmetry is broken. Consequently, instability mode grows in time, eventually leading to the emission of the vortex. The numerical duration of the simulation is insufficient to capture the impact of this phenomenon on the integrated quantities in Fig.~\ref{fig:energy_n=1}. However, we have already characterized it in~\cite{Dvali:2023qlk} when studying soliton mergers. Therein, we noted that a vortex localized in the merged configuration is unstable, therefore resulting in its eventual ejection. Examples of such dynamics can be found at the following \href{https://www.youtube.com/watch?v=t29WUvZM-io&ab_channel=MichaelZantedeschi}{URL1} and \href{https://www.youtube.com/watch?v=zorkQSYCliU&t=0s&ab_channel=MichaelZantedeschi}{URL2}. 

A different behaviour is observed in the first panel of the second row for $Q=Q_{\rm s}/4$. In this case, the first collapse is extremely violent, and one might naively expect the features analogous to the case $n=0$. Instead, only a negligible charge is emitted by the configuration, which is localized in a large ring around its center. The ring is, effectively, a slowly rotating (quasi-)oscillon. Eventually, due to the broken axial symmetry, for $t\gtrsim 120m^{-1}$, the system fragments into smaller $Q$-balls (a phenomenon known in the case of $U(1)$-symmetric $Q$-balls, see, e.g.,~\cite{Battye:2000qj,Kinach:2022jdx}), which cannot be visualized due to the $y$ cut of the plane performed in the figure. The phase keeps rotating between the fragmented $Q$-balls due to the angular momentum of the configuration.
Correspondingly, significant amount of energy and angular momentum are expelled from the configuration, as can be visualized at the following~\href{https://youtu.be/boDpRXJnT5E}{URL} and Fig.~\ref{fig:energy_n=1} show.

Analogous dynamics, eventually leading to fragmentation is observed also for vanishing charge (red line). Energy is discretely emitted from the configuration as the outer wall of the bubble squeezes the central vortex. These correspond to the three observed pulses in the fourth panel of Fig.~\ref{fig:energy_n=1}.
Characterization of both phenomena of fragmentation, as well as vortex ejection, are beyond the scope of this work.

\subsection*{Derivative interaction case}
\begin{figure*}[th!]
    \centering
    \includegraphics[width=1.02\textwidth]{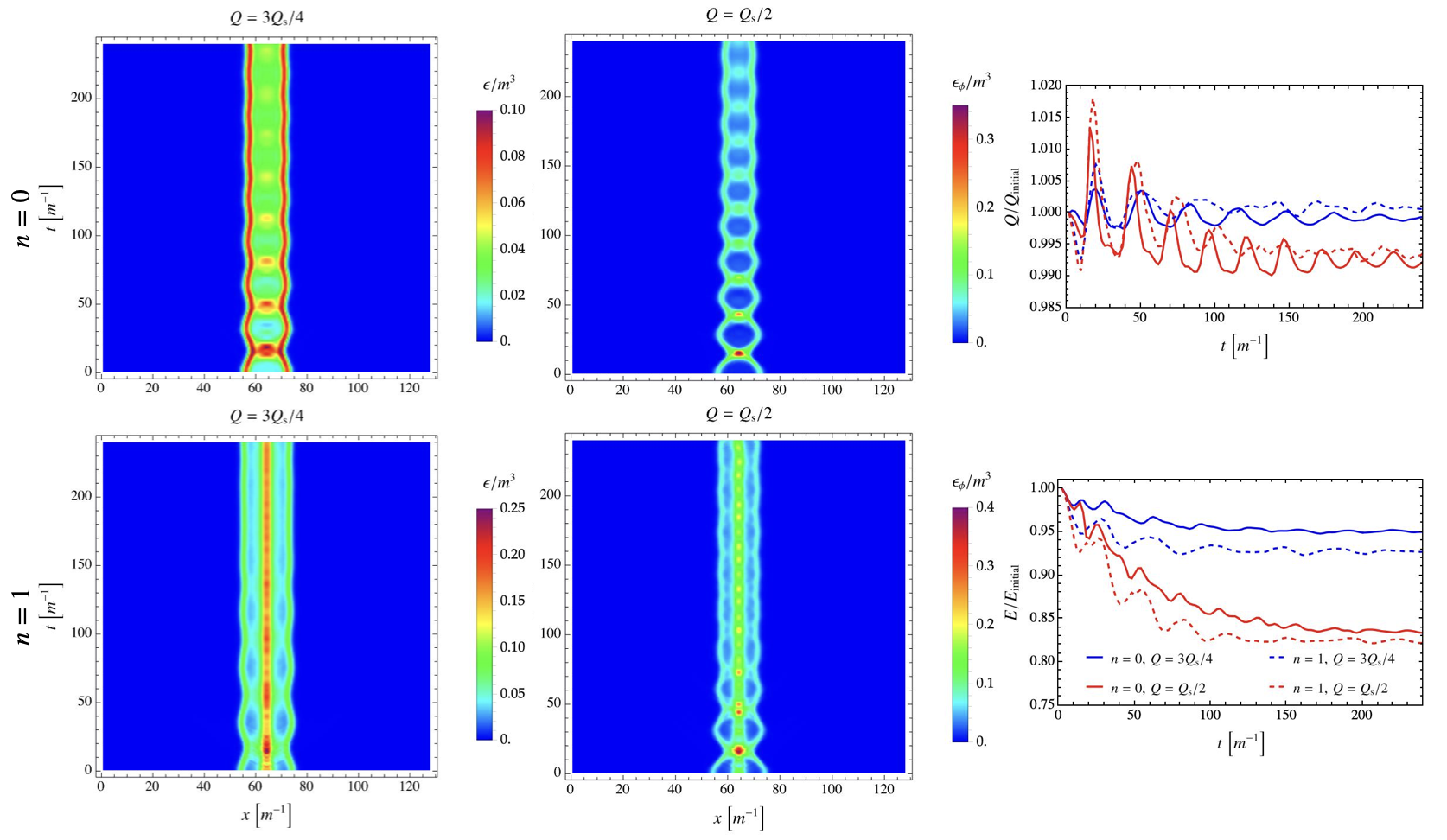}
    \caption{Energy density evolution for different initial charges for winding number $n=0$ (top row) and $n=1$ (bottom row) for non-vanishing interaction with derivatively coupled singlet $\chi$. Third column shows the integrated charge and energy as functions of time}
    \label{fig:energy_chicase}
\end{figure*}
In order to make the analogy between bubbles and black holes maximally transparent, and to allow for a significant relaxation of the configuration pulsations, we further extend the system by derivatively coupling the soliton field to a $SU(N)-$singlet scalar $\chi$ via the Lagrangian,
\begin{equation}
    \label{eq:chi}
    \mathcal{L}\supset \frac{1}{2}\partial_{\mu}\chi\, \partial^\mu\chi +\frac{g}{2}\chi \,{\rm Tr}(\partial_\mu \Phi)(\partial^\mu \Phi) - \frac{\lambda}{4} \chi^4\,.
\end{equation}
The above interaction term is sensitive only to the master mode of the saturon, while leaving the charge unaffected. The mechanism of stabilization by memory proceeds similarly to the previous cases. 

In units of $m$, the couplings are $g\simeq 0.5$ and $\lambda \simeq 0.15$ and we further initialize $\chi$ to be vanishing.
The resulting energy density evolution, adopting analogous initial conditions to the cases discussed in the previous Subsection, are displayed in Fig.~\ref{fig:energy_chicase}. 
The top row shows the dynamics for the case of winding number $n=0$, while the bottom corresponds to the case with vorticity. The cases with charge $Q=3Q_{\rm s}/4$ (first column) and $Q=Q_{\rm s}/2$ (second column) are sufficient for our discussion.

Although $\chi$ is massless, a full relaxation is not possible. In fact, the derivative interaction sources a tadpole for $\chi$, which therefore localizes on the bubble support.  A quartic coupling has been added in \eqref{eq:chi}, in order to tame the growth of $\chi$.  This setup effectively generates a local mass gap for $\chi$ of magnitude $\sim( \sqrt{g} \lambda\,\omega^2 \, f^2 )^{1/3}$. This backreacts on the capacity of the system to fully relax to a stationary configuration, and constitutes the main reason we observe oscillations - although much less pronounced - at late times. 
This is in stark contrast to the case without singlet $\chi$ of the previous Subsection cf., Figs.~\ref{fig:energy_n=0} and~\ref{fig:energy_n=1}, in which the bubble simply pulsates with almost constant frequency and amplitude. 

The integrated energy confirms this behaviour as shown in the third column of Fig.~\ref{fig:energy_chicase}. Clearly, significantly more energy is emitted in the case of initial charge $Q=Q_{\rm s}/2$ (red lines) as this is further away from being stationary and, therefore, has more energy available for the emission. 
In general, for $n=1$ (dashed lines), the energy damping is both faster and more efficient, due to the extra gradient energy of the bubble background around the vortex region.
\begin{figure*}[th!]
    \centering
    \includegraphics[width=1.01\textwidth]{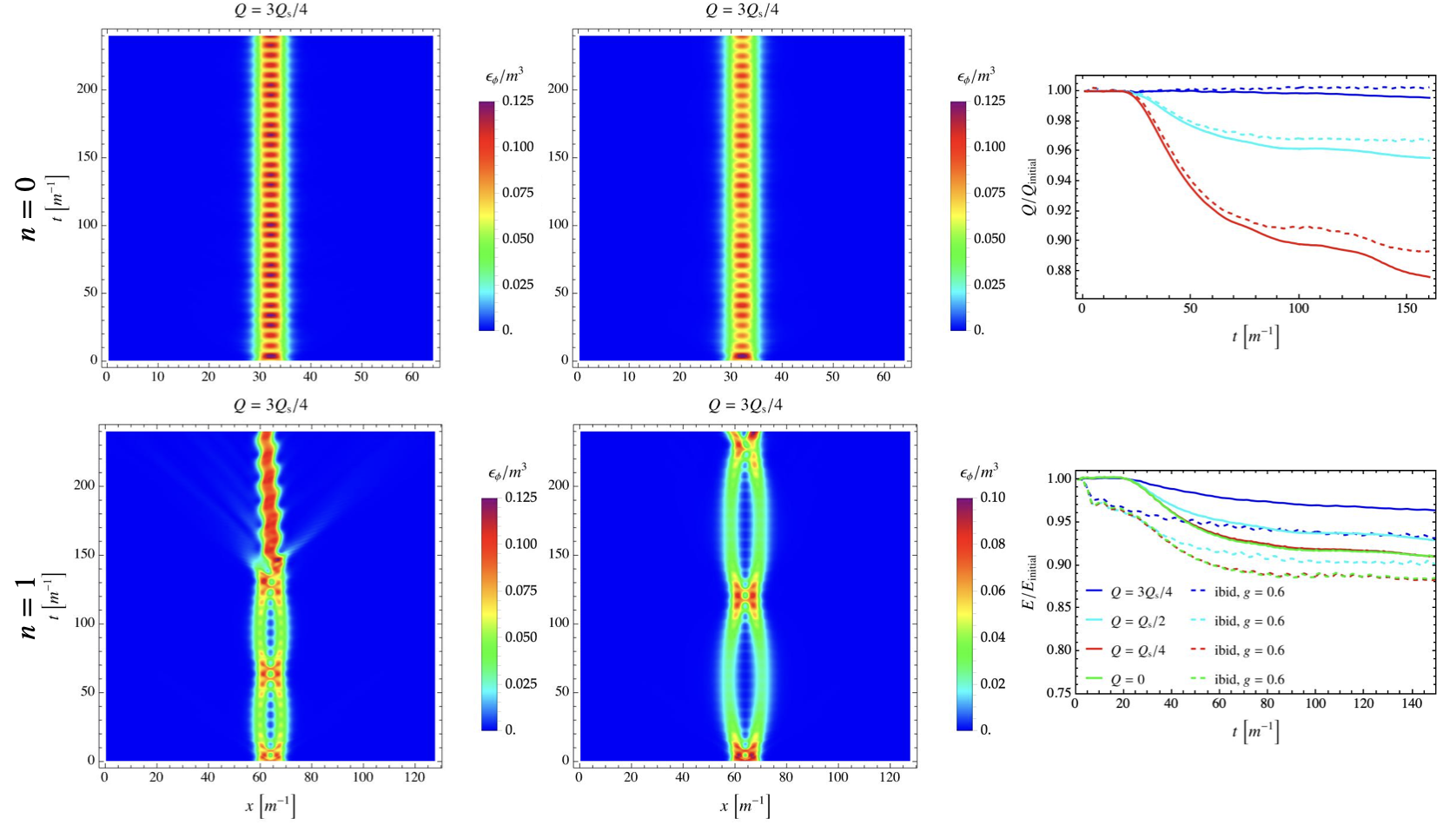}
    \caption{Energy density evolution for different initial charges for winding number $n=0$ (top row) and $n=1$ (bottom row) for vanishing (left column) and non-vanishing (central-column) interaction with derivatively coupled singlet $\chi$. The third column shows energy and charge evolution as functions of time for $n=0$ winding.}
    \label{fig:energy_thickwall_n0n1}
\end{figure*}
Nevertheless, charge is well-conserved - within $1\%$ - throughout the simulation time. This is expected given the nature of the coupling, not sensitive to the flavour of the bubble.

Compared to the case without $\chi$, the charge of the bubble is seen to be significantly more oscillatory. Notice that the derivative interaction leads to a redefinition of charge by a factor $1+g\chi$.  
In fact, the oscillatory behaviour is seen in correspondence of $\chi$ oscillations on the bubble support. The effect is larger at the beginning of the simulation since the $\chi$ field is very far from equilibrium and this leads to violent oscillations, modestly challenging the capabilities of our numerical implementation (as expected, given that we are dealing with a derivative interaction term)\footnote{
In order to be exactly charge and energy conserving, an indirect inversion method should be used when computing the time-evolution involving a matrix of both $\phi$ and $\chi$ at each lattice point. This is not very efficient, as a huge matrix should be inverted at once, leading to a significant restriction on the lattice dimension.
Instead, we adopt a direct method, based on finite difference, and compute each $t+1$ component of the field restricting to neighbouring points from time slices at $t$ and $t-1$ - $t$ being the index characterizing the step number. In practice, an exact method for the $\phi$ field evolution is used, evaluating all time derivatives with central scheme at the time-slice $t$ and using it to obtain the $t+1$ component. However, then to  update the $\chi$ field it becomes necessary to use a backward derivative for source term proportional to $g$. This induces an error in our algorithm dependent on the time-step size which, we verified, reflects - partly - the amplitudes of the peaks in the first panel of the third column of Fig.~\ref{fig:energy_chicase}.}. 

Finally, analogously to the case of the previous Subsection, in the last panel we can observe the onset of the vortex ejection phenomenon towards the final stage of the evolution.

\subsection*{Thick Wall regime} \label{subsec:thickwall}

 So far, our numerical analysis focused on bubbles in the thin-wall regime.
Although such bubbles are undersaturated in terms of the bound 
(\ref{AreaSbound}), they nevertheless represent the systems of enhanced 
capacity of information storage. 
This is fully sufficient for understanding the key features of the memory burden effect and mapping it on black holes. In fact, the 
thin-wall bubbles are very well suited for 
capturing the type-$II$ nature of the black hole memory burden
effect.

On the other hand, in the thick-wall regime the bubble/$Q$-ball represents a \textit{saturon} (see discussion around Eq.~\eqref{eq:numberofstates}).  This has advantage of reproducing the feature of a black hole in terms 
of the area-law scaling of entropy (\ref{AreaSbound}). 
Therefore, we shall also study this case. 

 As we will see, at a qualitative level, statements analogous to the ones of the previous Subsection hold also in this regime, providing further evidence for the generality of the stabilization by the burden of memory. 
 However, there are some quantitative differences, since the memory burden effect in thick-wall bubbles is of type-$I$. 

The two panels in the third column of Fig.~\ref{fig:energy_thickwall_n0n1} show the energy and the charge of the configuration as functions of time for the four cases of underburdened bubbles, corresponding to the four different colors and compare the cases without (continuous) and with (dashed line) the derivatively-coupled singlet $\chi$. For simplicity, we focus on the case of zero winding. Moreover, we chose $\omega_{\rm s}= 0.8\, m$, while keeping other parameters unchanged.

One notable feature in the thick wall regime is that the constituents within the bubble have larger frequency. Therefore, it is 
easier to excite modes above the asymptotic mass gap, as compared to thin-wall case. This, in turn, leads to a progressively larger relative emission of charged particles as the initial charge of the configuration decreases. This is due to the fact that the system is in the type-$I$ regime, which, for larger $\omega$, has a worse storing efficiency - see Eq.~\eqref{EpsB}. 

Remarkably, charge emission seems alleviated in presence of $\chi$-field, as another channel becomes available for relaxation, allowing the system to reach equilibrium while retaining information. On the other hand, a significant amount of pulsation energy is depleted due to the presence of the singlet.  

A further comment related to the behaviour of the system at small 
and vanishing charge (green and red line respectively) is due. In this case, the system does not collapse, as opposed to the thin wall scenario. Instead, only a small fraction of the initial energy is emitted; the reason being that the ``oscillon'' configuration is not sufficiently excited to produce asymptotic quanta. This basically results in stationary configurations. Equally stated, the oscillation frequency of the 
oscillon is smaller than the asymptotic mass gap. 

For completeness, we report in Fig.~\ref{fig:energy_thickwall_n0n1} also the energy density evolution of underbudened bubbles in the thick-wall regime for the winding numbers $n=0$ (first row) and $n=1$ (bottom row). In the left column the singlet coupling is $g=0$, while in the central column $g=0.6$. In the $n=1$ case, we can observe a marginal energetic emission due to the presence of the singlet $\chi$ channel of about $5\%$-$10\%$ of its total energy. 

Noticeably, in the third panel, a vortex ejection takes place at $t\simeq 150 m^{-1}$. Correspondingly, the configuration looses about $40\%$ ($30\%$) of its initial energy (charge). Moreover, since the vortex is responsible for the spin, its ejection causes a drop of about $90\%$ of the total angular momentum (we, once gain, refer the reader to~\cite{Dvali:2023qlk} for a detailed discussion of this phenomenon). In the case of derivative interaction (central panel) the ejection takes place later, towards the end of the simulation time, indicating that the derivative coupling can stabilize the vortex within the $Q$-ball support.  

\section{Implications and Outlook}\label{sec:implications}

The goal of the present paper was to further investigate the physical nature and the extent of universality of the memory burden phenomenon~\cite{Dvali:2018xpy, Dvali:2020wft}. 
After introducing the essence and generic features of the phenomenon, we have studied its manifestations in solitons building up on the previous work~\cite{Dvali:2021tez}. 

We have established a close correspondence with the expected features of the memory burden phenomenon in black holes. In this analysis, we took a double approach. 
On one hand, following~\cite{Dvali:2018xpy, Dvali:2020wft}, 
we have modelled the memory burden 
effect in black holes relying exclusively on their well-established features. These features unambiguously 
place black holes in the category of objects of enhanced capacity 
of information storage susceptible to the memory burden effect.   

  On the other hand, we used a microscopic theory of black hole's quantum $N$-portrait~\cite{Dvali:2011aa, Dvali:2012en} as the reference point for the consistency checks of our results.  
  
The emerging picture speaks in favor of close similarities between black holes and other objects of enhanced capacity of information storage, such as the solitonic saturons~\cite{Dvali:2019jjw, Dvali:2019ulr, Dvali:2020wqi,  Dvali:2021jto,  Dvali:2021ooc, Dvali:2021rlf,  Dvali:2021tez, Dvali:2021ofp, Dvali:2023qlk}.   
 
We have confronted the stabilization via memory burden with the stabilization by means of the classical hair and outlined the fundamental differences between the two mechanisms.   

 We have predicted that the memory burden effect 
 induces the mass splitting between initially-degenerate
 black holes according to differences in their information patterns.  

We have performed numerical analysis for various regimes of the solitonic memory burden, with or without the topological winding numbers. The results of these numerical simulations 
fully confirm our analytic conclusions.  

Let us now briefly go over some implications of the memory burden effect and the future prospects. \\

\subsection*{New dark matter window for PBH }

As put forward in~\cite{Dvali:2020wft}, one immediate implication for black hole stabilization by the memory burden effect is the opening of a new window for PBH dark matter in the range of masses below $\sim 10^{14}$g.  
The idea of PBH goes back to~\cite{Zeldovich:1967lct,Hawking:1971ei,Carr:1974nx}, and the proposal of PBH composition of dark matter is also old~\cite{Chapline:1975ojl} (for a review, see~\cite{Carr:2016drx}).
However, in the standard treatment, in 
which one extrapolates the Hawking 
regime till the very end of black hole existence, 
the PBH of masses below $\sim 10^{14}$g
were assumed to be  excluded from dark matter, 
since they were expected not to survive till the current epoch. 

The understanding that black holes undergo the memory burden effect~\cite{Dvali:2018xpy} and can get stabilized by it~\cite{Dvali:2020wft}, drastically changes the 
standard view. PBHs in the 
wide range of masses below $10^{14}$g can now account for the entirety of dark matter. The proof-of-concept type examples of light PBHs (e.g., with masses $\sim 10^8$g) 
which satisfy all the known constraints, were originally given in~\cite{Dvali:2020wft}.
      
Furthermore, the possibility of dark matter in 
the form of PBHs stabilized by memory burden in the mass range above $\sim 10^4\rm g$  (corresponding to most conservative 
value $k =1$ in  (\ref{tauBHk}))  was discussed in~\cite{Dvali:2021byy}. This paper proposes an explicit cosmological mechanism for the formation of such PBHs in the right abundance.    
   
A further analysis of various constraints on light ($ \lesssim 10^4\rm g$) PBH dark matter stabilized  by memory burden was offered in two recent papers~\cite{Alexandre:2024nuo, Thoss:2024hsr}.  
In particular, it was shown that PBHs that enter the memory burden phase prior to BBN epoch, are unconstrained from BBN and CMB. For the  most conservative estimate of 
the post memory-burden lifetime, corresponding to $k=1$ in (\ref{tauBH}), 
these are PBHs with masses below $10^9$g. Such PBHs can easily compose the entirety of dark matter.
Further implications of the memory burden effect for PBHs 
can be found in~\cite{Balaji:2024hpu, Haque:2024eyh} (see~\cite{Riotto:2024ayo} for some future prospects).
  
The analysis of the present paper predicts a spread in PBH dark matter masses, regardless of their production mechanism.  
This is due to the mass-splitting of the remnants evolving from initially-degenerate black hole states, as shown in equation (\ref{DMBHMBH}) and schematically 
described by Fig.~\ref{fig:schematic_memory}. 

Let us consider a black hole, formed at time $t_{\rm f}$
with initial mass $M_{\rm BH}(t_{\rm f})$,
which 
entered the memory burden phase around some time 
$t_{\rm M}$, prior to today's Hubble time. For definiteness, 
let us assume that the subsequent change of the mass is negligible.
For instance, this will be the case for $k > 0$
and $M_{\rm BH}(t_{\rm f}) \gtrsim 10^4$g. 

However, the mass of PBH at stabilization is determined by the 
memory burden that it carries (\ref{SpreadDM}). 
To the leading order, the burden is controlled by the occupation number of the memory modes, $N_{\rm G}$. 
As we have discussed, statistically, the most probable value of this number is of the order of the PBH initial entropy $S_{\rm BH}$.
The probability of forming a black hole
with memory pattern $N_{\rm G} \ll S_{\rm BH}$ is exponentially 
small. This implies that most of the PBHs that we 
observe today as dark matter carry the memory burdens of order the initial entropy $N_{\rm G} \sim S_{\rm BH}$, with the 
statistical spread given by (\ref{PofNG}). 
This translates as the corresponding spread in PBH masses via
(\ref{SpreadDM}). 

In particular, PBHs of masses $M_{\rm BH} \lesssim  10^{14} {\rm g}$
are subjected to this spread. 
Notice that  
in case of existence of a large number of hidden particle species,
the time $t_{\rm M}$ gets shortened as 
(\ref{tMsp})~\cite{Dvali:2021bsy}. Due to this, the memory burden 
can affect much heavier PBH dark matter~\cite{Alexandre:2024nuo}. 
Correspondingly, the prediction of the mass-spread will extend 
to such black holes. 

Notice that the spread is predicted even   
if the production of PBHs is sharply peaked at  
a particular mass.

\begin{figure}
\includegraphics[width=.45\textwidth]{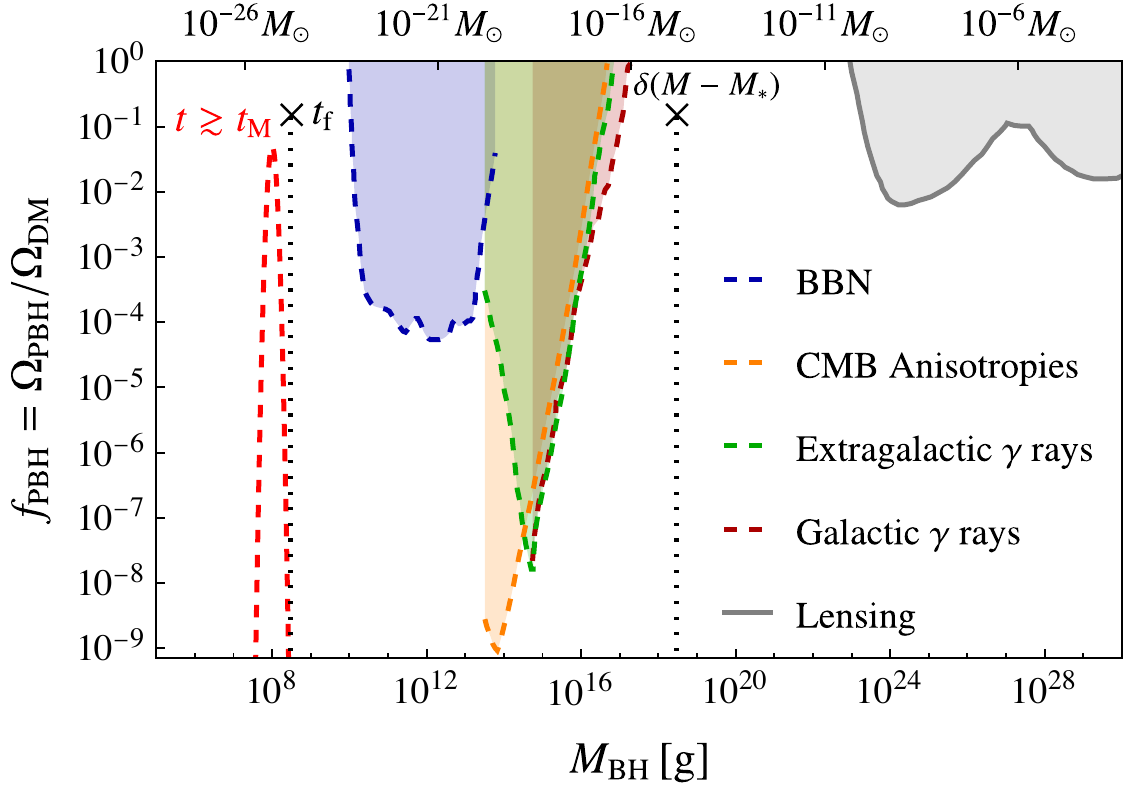}
\caption{$f_{\rm PBH}$ as a function of $M_{\rm BH}$. Shaded areas represent existing constraints - we refer the reader to~\cite{Carr:2009jm,Auffinger:2022khh,Thoss:2024hsr} for an accurate description. The ones on the left follows from Hawking evaporation and are, therefore, only mildly affected by the most conservative estimates of memory burden effect, $k=1$ [see~\eqref{tauBH}]. Dotted lines represent the monochromatic distribution at formation time $t_{\rm f}$. PBHs lighter than $10^{14}$g (ignoring corrections due to large number of species) are affected by the burden over cosmological times, resulting in a smearing of the distribution, with spread of order $M_{*}$, for $t\gtrsim t_{\rm M}$. The qualitative spread is schematically shown by the red-dashed line. }
\label{fig:fpbh}
\end{figure}

A schematic representation of this effect is provided in Fig.~\ref{fig:fpbh}, where the fraction of PBH energy density as dark matter, $f_{\rm PBH}$, is shown as a function of the PBH mass, $M_{\rm BH}$. The shaded areas denote the existing bounds. To explicate our point, at formation time, $t_{\rm f}$, we consider two monochromatic distributions peaked at $M_* \sim 10^{18}$g and $10^8$g.

The former is in the currently unconstrained asteroid mass window where PBHs can constitute an $\mathcal{O}(1)$ fraction of dark matter.
Ignoring a possibility of large number of hidden particle species (\ref{tMsp}), these objects are unaffected by their memory over the cosmological timescales.

The latter (lighter) PBH distribution, instead, lies in the new mass window opened by the memory burden effect. At late times, $t\gtrsim t_{\rm M}$, PBHs are stabilized by their memory. Consequently, the distribution - pictorially represented by the red-dashed line - develops a spread of order $M_{*}$ according to \eqref{SpreadDM}.

Of course, the quantitative confrontation 
of the model-independent spread (\ref{SpreadDM}) with the 
model-dependent one coming from particularities of a production mechanism 
can only be performed within a specific cosmological scenario 
providing such a mechanism. For example, the mechanism of~\cite{Dvali:2021byy} is based on black hole production due to collapse of confining 
strings connecting the heavy quarks and antiquarks. Strings are produced and stretched during inflation. After reentering the Hubble 
patch, the string tension pulls the quark and antiquark towards 
each other, colliding them and forming a PBH (see~\cite{Dvali:2022vwh} for a numerical simulation of the dynamics).  

This mechanism has certain intrinsic spread of PBH masses, due to the factors such as the duration of inflation. 
The memory burden imposes an additional spread (\ref{SpreadDM}), due to a 
statistical distribution of the initial memory burden of 
the type (\ref{PofNG}).  

We must note that possibility of black hole stabilization at macroscopic size due to quantum backreaction has been suggested previously in~\cite{Dvali:2012rt}.  
Even if such additional mechanisms are realized, the memory 
burden effect must be taken into account regardless, since  
it represents a generic physical mechanism of stabilization.
Most importantly,   
this phenomenon is universal for 
systems of enhanced information capacity  
and independent on particularities of a microscopic picture~\cite{Dvali:2018xpy,Dvali:2020wft}~\footnote{
In a separate context, the memory burden effect can play 
an important role in dark matter composed of non-gravitational saturons~\cite{Dvali:2023xfz}.}.

\subsection*{Implications for Inflationary Cosmology}

As proposed in~\cite{Dvali:2018ytn,Dvali:2021bsy},
the memory burden effect can be applicable to  
cosmological space-times such as de Sitter or inflation.  
This is due to the fact that, just like a black hole, a de Sitter Hubble patch of 
radius $R$ represents 
a system of enhanced information capacity. The evidence for this is provided  by Gibbons-Hawking entropy of de Sitter space $S_{\rm GH}$~\cite{Gibbons:1977mu}, which is very similar to the 
Bekenstein-Hawking entropy of a black hole (\ref{SBH}).  

In fact, treated as a coherent state of gravitons constructed 
on top of the Minkowski  vacuum,  the quantum portrait of 
de Sitter~\cite{Dvali:2013eja, Dvali:2014gua, Dvali:2017eba, Berezhiani:2021zst} shares some key features 
with a similar portrait of a black hole~\cite{Dvali:2011aa}. 
For example, the Gibbons-Hawking radiation is a result of 
the quantum depletion of the graviton coherent state.  

It was suggested in~\cite{Dvali:2013eja, Dvali:2014gua, Dvali:2017eba}
 that the 
back-reaction from this radiation necessarily leads 
to a gradual loss of coherence and, subsequently, to  a
complete breakdown of the classical description. 
This happens after a so-called {\it quantum break-time}, $t_{\rm Q}$.

The concept of quantum break-time was originally introduced in~\cite{Dvali:2013vxa} in the study of quantum evolution 
in a prototype many-body model invoked  in~\cite{Dvali:2012en} 
as a toy analog of a black hole $N$-portrait~\cite{Dvali:2011aa}.
 It was shown~\cite{Dvali:2013vxa} that in the regime in which the system possesses a
 Lyapunov exponent, $\lambda$, the quantum break-time can be logarithmically 
 short $t_{\rm Q} \sim \lambda^{-1} \ln (N)$, where 
 $N$ is the occupation number of constituents (in the present 
 language, a ``master mode"). 
   The same system, in the classically-stable regime exhibits 
  a power-law break-time, $t_{\rm Q} \propto \sqrt{N}$~\cite{Dvali:2015wca}.  
 
 The concept of quantum break-time was applied to de Sitter in~\cite{Dvali:2013eja, Dvali:2014gua, Dvali:2017eba} and  
 it was argued that for a cosmological constant source the 
 quantum break-time is bounded from above by $t_{\rm Q} \sim S_{\rm GH} R$. However, for a generic source, a shorter time-scale,  
 $t_{\rm Q} \sim R \ln (S_{\rm GH})$, emerges.  

  In recent years, other aspects of the phenomenon of quantum breaking have been studied extensively in various setups~\cite{Dvali:2013lva,Berezhiani:2016grw,Dvali:2017eba,Vachaspati:2017jtw,Dvali:2017ruz,Kovtun:2020ndc,Kovtun:2020udn,Berezhiani:2020pbv,Berezhiani:2021gph, Kaikov:2022sch, Michel:2023ydf,Berezhiani:2023uwt}.

Now, based on the universality of the memory burden effect~\cite{Dvali:2018xpy}, it was argued in~\cite{Dvali:2018ytn}
that the phenomenon must be operative in de Sitter and 
must contribute into quantum breaking. 

The idea of~\cite{Dvali:2018ytn} is that de Sitter must posses memory modes
which are responsible for initially-degenerate microstates
accounted by the Gibbons-Hawking entropy.   
It was concluded 
that due to the depletion of the graviton coherent state via Gibbons-Hawking radiation~\cite{Dvali:2013eja, Dvali:2014gua, Dvali:2017eba}, the memory burden effect must set in, 
latest, by the time $t_{\rm M} \sim S_{\rm GH} R$. This claim matches the previously suggested limit on quantum break-time of de Sitter due to the loss of coherence and self-entanglement~\cite{Dvali:2013eja, Dvali:2014gua, Dvali:2017eba}.  
It is also very similar to the upper bound (\ref{tMMM}) on the memory burden time of a black hole.  
I general, the memory burden effect is one of the main engines      
of quantum breaking of de Sitter~\cite{Dvali:2018ytn, Dvali:2021bsy}.

The observational signatures of the Hubble memory burden effect still remain to be understood. In inflationary context, it is expected to create departures from the standard semi-classical  spectrum of density perturbations, with the amplitude 
increasing with the duration of inflation~\cite{Dvali:2018ytn, Dvali:2021bsy}. 
In this way, the memory burden effect provides a quantum clock that records the entire duration of the inflationary phase. 

Although the analysis of the present paper was not directly 
intended at de Sitter, it nevertheless supports the 
generic expected features previously proposed in~\cite{Dvali:2018ytn, Dvali:2021bsy}. 
This motivates future studies of applications of the memory burden effect to various cosmological backgrounds.  

\subsection*{Memory Burden in the Standard Model?}

 It was argued recently~\cite{Dvali:2021ooc} that 
 the Standard Model contains 
 a saturon in form of a color glass condensate 
 (CGC)~\cite{Gelis:2010nm}. This substance represents a saturated state of $N_{\rm glue} = 1/\alpha_{\rm s}(Q_{\rm s})$ gluons
 where $\alpha_{\rm s}(Q_{\rm s})$ is a running  QCD coupling evaluated at 
 a saturation scale $Q_{\rm s}$. As discussed in~\cite{Dvali:2021ooc}, CGC exhibits a striking correspondence with the black hole $N$-portrait~\cite{Dvali:2011aa}, with the gluons of CGC mapped on the graviton constituents of a black hole. 
 
  Being a saturated state,
 CGC is expected to be subjected to the memory burden effect,
 with potentially observable consequences. In particular, 
 this can be manifested in emission of quanta by a factor 
 of $\alpha_{\rm s}$ softer than the saturation scale $Q_{\rm s}$~\cite{Dvali:2021jto}.  \\
 
\paragraph*{\textbf{Acknowledgements.} }
 The work of G.D. was supported in part by the Humboldt Foundation under the Humboldt Professorship Award, by the European Research Council Gravities Horizon Grant AO number: 850 173-6, by the Deutsche Forschungsgemeinschaft (DFG, German Research Foundation) under Germany’s Excellence Strategy - EXC-2111 - 390814868, Germany’s Excellence Strategy under Excellence Cluster Origins 
EXC 2094 – 390783311. 

J.S.V.B. acknowledges the support from the Departament de Recerca i Universitats from Generalitat de Catalunya to the Grup de Recerca ‘Grup de Fisica Teorica UAB/IFAE’ (Codi: 2021 SGR 00649) and the Spanish Ministry of Science and Innovation (PID2020-115845GB- I00/AEI/10.13039/501100011033). IFAE is partially funded by the CERCA program of the Generalitat de Catalunya. This study was supported by MICIIN with funding from European Union NextGenerationEU(PRTR-C17.I1) and by Generalitat de Catalunya

M.Z. acknowledges support by the National Natural Science Foundation of China (NSFC) through the
grant No. 12350610240 “Astrophysical Axion Laboratories”. \\

Disclaimer: Funded by the European Union. Views and opinions expressed are however those of the authors only and do not necessarily reflect those of the European Union or European Research Council. Neither the European Union nor the granting authority can be held responsible for them.\\

\bibliography{biblio}
\end{document}